\documentclass[12pt]{article}
\usepackage[margin=1.25in]{geometry}
\usepackage{mathrsfs, mathtools}
\usepackage[title,toc,titletoc]{appendix}
\usepackage{titlesec}
\titleformat{\section}[block]{\large\normalfont\filcenter}{\thesection.}{.5em}{}
\titleformat{\subsection}[runin]{\normalfont\bfseries\filright}{\thesubsection.}{.5em}{}
\usepackage[dvipsnames]{xcolor}
\usepackage{bbm}
\usepackage{bm}
\usepackage{amsmath, amssymb}	
\usepackage[shortlabels]{enumitem}
\usepackage[hang,flushmargin]{footmisc}
\usepackage[round]{natbib}
\usepackage[colorlinks=true,linkcolor = blue, citecolor=blue, hyperfootnotes=false, hyperindex,breaklinks]{hyperref}
\usepackage{cleveref}
\usepackage{bigints}

\usepackage{pgf}
\usepackage[justification=centering]{caption}
\usepackage{subcaption}
\usepackage{tikz}
\usepackage{pgfplots}
\pgfplotsset{compat=1.18}
\usetikzlibrary{patterns, decorations.pathreplacing}
\usetikzlibrary{spy}
\usepackage{hhline}
\usepackage{float}
\usepackage{siunitx}
\usepackage{textcomp}
\usepackage{graphics}
\usepackage{graphicx}

\newtheorem{proposition}{Proposition}

\newtheorem{corollary}{Corollary}

\newenvironment{proof}[1][Proof]{\textbf{#1.} }{\  \rule{0.5em}{0.5em}}

\newcommand{\ts}{{\theta^{\vphantom{X}^*}\hspace{-.1em}}}
\newcommand{\tc}{{\theta^{\vphantom{X}^c}\!}}

\usepackage{setspace}
\usepackage{multirow}
\definecolor{shade}{gray}{.7}
\usepackage{xr}
\makeatletter
\newcommand*{\addFileDependency}[1]{
  \typeout{(#1)}
  \@addtofilelist{#1}
  \IfFileExists{#1}{}{\typeout{No file #1.}}
}
\makeatother

\begin{document}
\title{Luck Out or Outpay? \\Competing with a Public Option\thanks{I am grateful to Arjada Bardhi, Zi Yang Kang, Erik Madsen, Ellen Muir, Bobak Pakzad-Hurson, Anne-Katrin Roesler, Alex  Teytelboym, Michael Thaler, Kun Zhang, and anonymous reviewers for insightful discussions and comments. I have also benefited tremendously from the feedback of the co-editor and two anonymous referees.}}

\author{Teddy Mekonnen\thanks{Department of Economics, Brown University. Contact: \href{mailto:mekonnen@brown.edu}{mekonnen@brown.edu}}}
\date{\today}

\maketitle
\thispagestyle{empty}
\setcounter{page}{0}
\begin{abstract}
This paper analyzes the strategic interactions between a profit-maximizing monopolist and a free, capacity-constrained public option. By restricting its own supply, the monopolist intentionally congests the public option and induces rationing, which increases consumers' willingness to pay for guaranteed access. Counterintuitively, expanding the public option's capacity may raise the monopoly price and lower consumer welfare. I derive conditions under which all buyer types benefit from a capacity expansion, and extend these results to a setting where an oligopoly competes with a public option. These findings have implications for mixed public-private markets, such as housing, education, and healthcare.
\end{abstract}

\newpage
 
 \begingroup
\allowdisplaybreaks

\section{Introduction}\label{intro}
In the housing, education, and healthcare sectors, government-run programs provide a critical safety net by making essential goods and services accessible through free or heavily subsidized offerings. Prominent examples include subsidized public housing in Singapore, tuition-free universities in Norway and Germany, and free healthcare in the United Kingdom, Canada, and Australia. These programs, however, often face demand that exceeds their supply, leading to rationing through waiting times or lotteries. At the same time, they frequently operate alongside profit-maximizing firms that provide the same goods and services as either a supplement or an alternative to the rationed public provision. Thus, such settings are best understood as \emph{mixed markets} comprising a private market and a capacity-constrained \emph{public option}. 

In this paper, I focus on the case in which the private market and the public option are substitutes for one another. For instance, individuals often choose between public and private education, but rarely combine the two. In such settings, public options serve not only as a safety net, but also as potential sources of competitive pressure on their private counterparts. This paper examines the strategic interactions that emerge when private providers compete against a public option. In particular, can the government resolve the frequently observed rationing at a public option by simply increasing its capacity? And are consumers better off in a mixed market than they would be in a market served only by private providers, or only by a public option?

The central insight of the paper is that rationing at a public option is an equilibrium outcome that is shaped by firms in the private market. By restricting their own supply, firms redirect demand toward the public option. This worsens rationing at the public option and raises consumers' willingness to pay for guaranteed private access. This strategic response implies that expanding the capacity of a public option may have unintended negative consequences: it can lead to higher prices in the private market and lower aggregate consumer welfare. The paper characterizes when these adverse price effects arise, when capacity expansions benefit all consumers, and how the equilibrium changes with the degree of competition among private providers and with the timing of access to the public option.

To that end, I begin with a baseline model of a mixed market in which the same good is supplied by a capacity-constrained public option and a profit-maximizing monopolist. There is a unit mass of risk-neutral consumers with heterogeneous valuations for the good. Each consumer has unit demand and chooses between the two suppliers. The public option offers the good for free but randomly rations it whenever demand exceeds capacity. In contrast, the monopolist faces no capacity constraints and sells the good to maximize its profit. Within this parsimonious model, I characterize the mixed market equilibrium outcome and study its welfare implications. 

To build some intuition, consider the following example that illustrates the paper's main unintended-consequences result: a public option introduced as a safety net can lead to a congested public option and a higher private-market price. There is a unit mass of buyers: a mass $1/4$ have a low valuation $\theta=1$ (type $L$), a mass $1/2$ have a valuation $\theta=4$ (type $M$), and a mass $1/4$ have a high valuation $\theta=11$ (type $H$). Initially, the buyers are served only by a monopolist with zero marginal cost. The profit-maximizing price in this baseline is $p^M=4$, which yields a profit of $3$.\footnote{For comparison, a price of $p=1$ yields a profit of $1$, and a price of $p=11$ yields a profit of $2.75$.} Thus, all type-$L$ buyers are excluded from the market and the type-$M$ buyers purchase from the monopolist but earn no surplus. Hence, the aggregate consumer surplus is simply the surplus of the type-$H$ buyers, which equals $7/4$.

Next, suppose a well-intentioned planner introduces a free public option with capacity $k=1/4$ as a safety net for the excluded type-$L$ buyers. If, as the planner naively assumes, only the targeted type-$L$ buyers opt for the public option, there would be no rationing. All buyers would receive the good with probability one, so aggregate consumer surplus would rise from $7/4$ to 2.

However, this conclusion hinges on the assumption that neither the monopolist nor the consumers adjust their behavior in the presence of the public option. In reality, because the public option is free, it inevitably attracts the type-$M$ buyers who were previously purchasing from the monopolist but earning zero surplus. The monopolist thus faces a strategic pricing choice: compete with the public option to retain the mid-valuation buyers, or abandon them to sell only to the high-valuation buyers. 

If the monopolist chooses to compete for the type-$M$ buyers, its profit-maximizing price is $p^*=2<p^M$.\footnote{To see why, suppose a fraction $x\in [0,1]$ of the type-$M$ buyers opt for the public option. All type-$L$ buyers also opt for the public option while all type-$H$ buyers purchase from the monopolist. Then the public option allocates a good with probability $1/(2x+1)$. The monopoly price that leaves the type-$M$ buyers indifferent between the public option and the private market becomes $P(x)=8x/(2x+1)$. The monopolist's profit is maximized at $x^*=1/2$, which induces a price $p^*=P(x^*)=2$.} Half of the type-$M$ buyers and all of the type-$H$ buyers purchase from the monopolist while the remaining half of the type-$M$ buyers and all of the type-$L$ buyers rely on the public option. As a result, demand at the public option becomes $d=1/2$, exceeding the capacity. The public option is thus rationed; it allocates the good with probability $k/d=1/2$. Importantly, the price of $p^*=2$ leaves the mid-valuation buyers indifferent between the private market and the public option, and yields the monopolist a profit of 1.  

In contrast, if the monopolist abandons the type-$M$ buyers entirely, its profit-maximizing price is $p^{**}=22/3>p^M$. All the type-$H$ buyers purchase from the monopolist while all the type-$L$ and type-$M$ buyers rely on the public option. As a result, demand at the public option surges even more to $d=3/4$, and the public option allocates the good with probability $k/d=1/3$. Once again, observe that the price $p^{**}$ leaves the high-valuation buyers exactly indifferent between the private market and a public option, and yields a profit of $11/6$. 

Consequently, the monopolist strictly prefers to abandon the mid-valuation buyers. Much to the dismay of the naive planner, the monopolist raises its price following the introduction of the public option and the market outcome ultimately forces buyers to choose between a heavily congested public option and a significantly more expensive private market. Aggregate consumer surplus falls from $7/4$ in the monopoly-only baseline to $5/3$ following the introduction of the public option. Strikingly, consumers in this example would have been better off without the free, capacity-constrained public option.

\begin{figure}[htpb]
\centering
\begin{tikzpicture}[
    scale=1.0,
    every node/.style={font=\small},
    public/.style={fill=green!25, draw=black, line width=0.4pt},
    private/.style={fill=blue!20, draw=black, line width=0.4pt},
    excluded/.style={fill=gray!25, draw=black, line width=0.4pt},
    slicelabel/.style={font=\scriptsize, align=center},
    titlelabel/.style={font=\small\bfseries, align=center},
    notelabel/.style={font=\scriptsize, align=center}
]

\def\R{1.45}

\begin{scope}[shift={(0,0)}]
    \node[titlelabel] at (0,2.0) {Monopoly only};

    \draw[private] (0,0) -- (0:\R) arc[start angle=0, end angle=90, radius=\R] -- cycle;
    \draw[excluded] (0,0) -- (90:\R) arc[start angle=90, end angle=180, radius=\R] -- cycle;
    \draw[private] (0,0) -- (180:\R) arc[start angle=180, end angle=360, radius=\R] -- cycle;

    \node[slicelabel] at (45:0.88) {$\theta=11$};
    \node[slicelabel] at (135:0.88) {$\theta=1$};
    \node[slicelabel] at (270:0.78) {$\theta=4$};

    \node[notelabel, text width=3.2cm] at (0,-1.95) {$p^M=4$, CS $= 7/4$};
\end{scope}

\begin{scope}[shift={(4.4,0)}]
    \node[titlelabel] at (0,2.0) {Naive Planner};

    \draw[private] (0,0) -- (0:\R) arc[start angle=0, end angle=90, radius=\R] -- cycle;
    \draw[public] (0,0) -- (90:\R) arc[start angle=90, end angle=180, radius=\R] -- cycle;
    \draw[private] (0,0) -- (180:\R) arc[start angle=180, end angle=360, radius=\R] -- cycle;

    \node[slicelabel] at (45:0.88) {$\theta=11$};
    \node[slicelabel] at (135:0.88) {$\theta=1$};
    \node[slicelabel] at (270:0.78) {$\theta=4$};

    \node[notelabel, text width=3.2cm] at (0,-1.95) {$p^M=4$, CS $=2$};
\end{scope}

\begin{scope}[shift={(8.8,0)}]
    \node[titlelabel] at (0,2.0) {Equilibrium};

    \draw[private] (0,0) -- (0:\R) arc[start angle=0, end angle=90, radius=\R] -- cycle;
    \draw[public] (0,0) -- (90:\R) arc[start angle=90, end angle=180, radius=\R] -- cycle;
    \draw[public] (0,0) -- (180:\R) arc[start angle=180, end angle=360, radius=\R] -- cycle;

    \node[slicelabel] at (45:0.88) {$\theta=11$};
    \node[slicelabel] at (135:0.88) {$\theta=1$};
    \node[slicelabel] at (270:0.78) {$\theta=4$};

    \node[notelabel, text width=3.2cm] at (0,-1.95) {$p^{**}=22/3$, CS $=5/3$};
\end{scope}

\begin{scope}[shift={(1.65,-3.0)}]
    \draw[excluded] (0,0) rectangle +(0.35,0.25);
    \node[anchor=west, font=\scriptsize] at (0.45,0.125) {Excluded};

    \draw[public] (2.0,0) rectangle +(0.35,0.25);
    \node[anchor=west, font=\scriptsize] at (2.45,0.125) {Public option};

    \draw[private] (4.5,0) rectangle +(0.35,0.25);
    \node[anchor=west, font=\scriptsize] at (4.95,0.125) {Private market};
\end{scope}

\end{tikzpicture}
\captionsetup{oneside,margin={0cm,0cm},justification=justified, singlelinecheck=false}
\caption{Motivating example: Each pie slice represents the mass of a buyer type, while the color indicates whether that type is excluded, relies on the public option, or purchases from the private market. Each panel also notes the associated price and aggregate consumer surplus.}
\label{fig:motivating_example_pies}
\end{figure}

The motivating example cleanly highlights the strategic effects at play in a mixed market. The monopolist's pricing decision affects not only its own demand but also the public option's demand. The more the monopolist limits its own supply by excluding low-valuation buyers from the private market, the more congested the public option becomes. This induces rationing at the public option, making it a less appealing ``damaged good,'' which in turn allows the monopolist to extract higher rents from its remaining higher-valuation consumer base. As the example demonstrates, when this strategic effect is particularly pronounced, mixed markets feature higher prices and lower consumer welfare than a monopoly-only market.

More generally, I show that the unique equilibrium outcome of a mixed market always features rationing at the public option, regardless of its capacity. Formally, the unique equilibrium is characterized by a cutoff. Buyers whose valuation exceeds the cutoff purchase from the monopolist, while buyers whose valuation falls below the cutoff rely exclusively on the public option. The marginal buyer---whose valuation is exactly equal to the cutoff---is indifferent between guaranteed access to the good from the private market and a free but probabilistic allocation via the public option, where the allocation probability is itself endogenously determined by the mass of buyers with valuations below the cutoff.

Using this cutoff characterization, I analyze how the equilibrium outcome varies with the public option's capacity. This comparative statics analysis is particularly relevant for evaluating policies that introduce a public option into an initially monopolistic market, or expand the capacity of an existing public option. I first show that the cutoff increases with the capacity. In other words, the monopolist responds by further restricting its own supply in order to keep the public option rationed. Consequently, the monopolist's market share and its profit decline as capacity grows. 

The implications for consumers, however, are more nuanced. Absent any strategic response by the monopolist, a capacity increase would naturally ease rationing at the public option. Yet, the monopolist offsets this benefit by raising the cutoff for guaranteed access, which shifts even more consumers to the public option. Furthermore, as the cutoff increases, the marginal buyer becomes a higher-valuation type, potentially leading to an increase in her willingness to pay for guaranteed access. 

I show that even with the monopolist's offsetting response, an increase in capacity always reduces congestion at the public option. Hence, even if a government or planner cannot eliminate rationing by increasing capacity, such a policy still increases the allocation probability and, therefore, the welfare of lower-valuation buyers who rely on the public option. On the other hand, the same capacity expansion could lead the monopolist to raise its price, as demonstrated by the motivating example. Hence, capacity-constrained public options may fail to exert competitive pressure. Nonetheless, I identify necessary and sufficient conditions on the buyers' type distribution under which a capacity expansion always lowers the monopoly price, benefiting even the higher-valuation buyers who do not use the public option.

I also analyze the impact on consumer welfare when a monopolist enters a market initially served by a standalone public option. Here, I find that consumer welfare improves unambiguously. High-valuation buyers gain from the introduction of guaranteed access via the monopolist, while low-valuation buyers benefit from reduced congestion at the public option as some of its demand shifts toward the private market.
 
Finally, I explore several extensions of the baseline model to demonstrate the robustness of the main insights. First, I show that the qualitative results extend to a setting where the private market is an oligopoly. In particular, the private market once again withholds supply to induce rationing at the public option. However, such strategic incentives are attenuated in an oligopolistic mixed market relative to a monopolistic one, and they vanish in the limit as the number of firms grows large. Second, I consider the case where the monopolist operates as a complement to the public option, with consumers first trying their luck at the public option and ``topping up'' their demand in the private market. Unsurprisingly, the strategic effects highlighted in the baseline model are no longer present in this alternative setting. However, even though consumers are no longer forced to choose between the two suppliers, I derive conditions under which consumer surplus is actually higher in the baseline model. Finally, I contrast the competition between a monopoly and a public option with that of a dominant firm facing a competitive fringe, highlighting the role of the fringe's supply elasticity in shaping the similarities and differences between the two models.

\subsection*{Related Literature:}\label{sec:relatedlit}This paper relates to a large literature on redistribution through public provision of goods or in-kind transfers. \cite{nichols1982targeting} show that participation costs can be used to screen for higher-need individuals, while \cite{blackorby1988cash} demonstrate that in-kind transfers can be more effective screening devices than cash. \cite{weitzman1977price} shows that random allocation may outperform market allocation when the welfare criterion differs from utilitarianism, and \cite{che2013assigning} show that even under utilitarian objectives, random allocations with resale can be superior to competitive markets when agents face budget constraints. More recently, \cite{condorelli2013market}, \cite{dworczak2021redistribution}, and \cite{akbarpour2024economic, akbarpour2024redistributive} apply mechanism design to study redistribution under general welfare criteria. However, all these papers consider a public option or transfer program in isolation. In contrast, this paper analyzes a setting in which a public option and profit-maximizing firms coexist within a broader mixed market, with no single provider fully determining market outcomes.

In this regard, the most closely related papers are \cite{besley1991public, coate1994pecuniary, kang2023public, kang2024optimal}. \cite{besley1991public} show that agents self-select into in-kind transfer programs when a private market is available, and \cite{kang2024optimal} study the optimal design of such transfers in the presence of a competitive private market. However, both papers abstract from the impact of the transfer program on prices in the private market. \cite{coate1994pecuniary}, in contrast, examine how in-kind transfers affect pricing decisions in the private market, but treat participation in the program as exogenous. \cite{kang2023public} studies how the government's choices regarding the public option's pricing, quality, and allocation affect both the composition of demand in the private market and its price.

The central focus across these papers is the design of the public option in the presence of a perfectly competitive private market. By contrast, this paper considers the profit-maximization problem of firms with market power when they compete with a public option. In this regard, this paper is also related to the literature on the competition between a dominant firm and a secondary fringe market \citep{philippon2012optimal, tirole2012overcoming,calzolari2015exclusive}. While the public option in this paper also acts as a ``fringe'' by providing an outside option to consumers, it differs fundamentally from a traditional competitive fringe because both its capacity and price are fixed (with price equal to zero). 

The remainder of the paper proceeds as follows. I present the baseline model in \Cref{sec:Model}, the equilibrium characterization in \Cref{sec:equil}, and the main results in \Cref{sec:main}. \Cref{sec:extensions} discusses several extensions of the baseline model, and  \Cref{sec:conclusion} concludes. All proofs are contained in the \hyperref[sec:appendix]{Appendix}.

\section{Model}\label{sec:Model}

\subsection*{Setup:}
There is a unit mass of risk-neutral buyers, each with a unit demand for a good. A buyer's  valuation for the good, which is the buyer's private type, is denoted by $\theta\in\Theta\coloneqq [\underline \theta, \bar \theta]$, with $\bar \theta>\underline \theta\geq 0$. The type distribution in the market is denoted by $F$, which is assumed to be regular: it is continuously differentiable with a strictly positive density $f$ over $\Theta$, and its virtual value function 
\[
\varphi(\theta)\coloneqq \theta-\frac{1-F(\theta)}{f(\theta)}
\]
is strictly increasing.

Each buyer may acquire the good from one of two suppliers: a profit-maximizing monopolist or a public option. For simplicity, I assume that the monopolist faces no cost of production. If the monopolist sells the good to a type-$\theta$ buyer at price ${p}\geq 0$, the buyer earns $\theta-{p}$ while the monopolist earns ${p}$. 

The public option offers the good for free but is capacity-constrained: it can only supply a mass $k \in (0,1)$ of goods. When demand for the public option exceeds its capacity, it rations the good via a lottery. Specifically, if a mass $d\in[0,1]$ of buyers demands the public option, each receives the good with probability $\min\{1,k/d\}$, where $\min\{1,k/0\}=1$ by convention. In this case, a type-$\theta$ buyer's payoff from the public option is $\theta\cdot\min\{1,k/d\}$.

\subsection*{Timing:} The game proceeds as follows: First, the public option's capacity $k$ is established, and each buyer privately observes her type. While I treat $k$ as exogenous, the analysis extends to settings where a government optimally chooses $k$ (possibly at some cost) in this initial stage. Taking the capacity as given, the monopolist posts a price $p$ in order to maximize its profit.\footnote{The restriction to posted prices here is without loss of generality. \Cref{sec:post} considers the partial mechanism design problem of a monopolist when it competes with a public option and shows that a posted price is the profit-maximizing mechanism.} Each buyer then chooses between the public option and the monopolist  to maximize her payoff. Finally, buyers who choose the monopolist purchase the good at the posted price, while the public option allocates the good among those who chose not to purchase from the monopolist, using a lottery whenever demand exceeds capacity.\footnote{The timing is such that buyers are forced to pick one of the two suppliers. An equivalent formulation is that buyers first decide whether to purchase from the monopolist and turn to the public option only if they decline the monopolist’s offer. In \Cref{sec:complement}, I analyze an alternative timing in which buyers turn to the monopolist only if they were not lucky enough to receive the good for free from the public option. This alternate timing captures a setting in which the monopolist complements the public option, with buyers ``topping up'' their demand in the private market.}    

\section{Equilibrium}\label{sec:equil}
For the subsequent analysis, rather than considering a monopolist who posts a price, it is more convenient to consider a monopolist who chooses the quantity it supplies. To that end, consider a monopolist that supplies a mass $q\in[0,1]$ of the good. A price $P(q)$ clears the private market if exactly a mass $q$ of buyers choose to purchase the good at this price, while the remaining mass $1-q$ of consumers choose the public option. In this case, a type-$\theta$ buyer earns $\theta-P(q)$ if she purchases from the private market, and a payoff of $\theta\cdot\min\{1,k/(1-q)\}$ from the public option. Thus, type $\theta$  purchases from the monopolist if
\begin{align*}
\theta\left(1-\min\left\{1, \frac{k}{1-q}\right\}\right)\geq P(q).
\end{align*}

The net gain from purchasing from the monopolist rather than relying on the public option is weakly increasing in $\theta$. Thus, without loss of generality, I restrict attention to market outcomes characterized by a cutoff: there exists a type $\tc(q)$ such that a buyer purchases from the monopolist at price $P(q)$ if and only if $\theta \geq \tc(q)$. Because $P(q)$ clears the private market, the cutoff $\tc(q)$ must satisfy $1 - F(\tc(q)) = q$. Moreover, the price must render the cutoff type $\tc(q)$ exactly indifferent between purchasing from the monopolist and relying on the public option. In other words, $\tc(q)$ is the \emph{marginal buyer type} and the inverse demand function must be given by 
\begin{equation}
\label{eq:market clearing price}
P(q)=\left(1-\min\left\{1, \frac{k}{F(\tc(q))}\right\}\right)\tc(q).
\end{equation}

The monopolist's problem can therefore be written as $\max_{q\in[0,1]} P(q)q$, where $P(q)$ is the inverse demand function derived in \eqref{eq:market clearing price}. Notice that both the quantity $q$ and the inverse demand function $P(q)$ can be expressed as functions of the cutoff type $\tc(q)$. Thus, the profit-maximization problem can be equivalently stated as choosing an optimal cutoff type:
\begin{equation}
    \label{eq:profit2}
    \max_{\theta\in\Theta}\,\left(1-\min\left\{1, \frac{k}{F(\theta)}\right\}\right) \theta\big(1-F(\theta)\big).
\end{equation}
This profit-maximization problem generalizes the standard monopoly-only benchmark in which $k=0$, where the profit-maximizing cutoff (and price) is given by $\theta^M\coloneqq\min\{\theta\in\Theta:\varphi(\theta)\geq 0\}$ \citep{myerson1981optimal}.

\begin{proposition}
\label{prop:optimal}
The monopolist's profit maximization problem \eqref{eq:profit2} has a unique solution $\ts\in \Theta$, which solves 
\begin{equation}
\label{eq:foc}
\left(1-\frac{k}{F(\theta)}\right)\varphi(\theta)=\frac{k}{F(\theta)^2}\,\theta\big(1-F(\theta)\big).
\end{equation}
In particular, $\max\{\theta^M, F^{-1}(k)\}<\theta^*< \bar \theta$, and 
\begin{enumerate}[$(\alph*)$]
    \item The monopolist supplies a quantity $q^*=1-F(\theta^*)$ at price $P(q^*)$,
    \item Buyer types $\theta\geq \theta^*$ purchase from the monopolist, and 
    \item Buyer types $\theta< \theta^*$ rely on the public option.
\end{enumerate}
\end{proposition}

\autoref{prop:optimal} implies that the unique equilibrium outcome segments consumers into ``high-value" and ``low-value" buyers based on their type. The distinction between the two groups is determined by a cutoff $\ts\in \Theta$. High-value buyers (types $\theta>\ts $) purchase the good from the monopolist, whereas  low-value buyers (types $\theta<\ts $) rely on the public option. Because $\ts>F^{-1}(k)$, the cutoff ensures that the public option is rationed. Thus, high-value buyers pay for guaranteed  access to the good, while low-value buyers rely on the free public option but accept the risk of not obtaining the good. Furthermore, the monopolist chooses to restrict its supply beyond what it would under a standard monopoly screening problem, i.e.,  $\ts>\theta^M$.  

The optimal cutoff $\ts$ uniquely solves \eqref{eq:foc}, which captures the monopolist's tradeoff from a marginal increase in the cutoff type. The left-hand side captures the marginal cost to the monopolist: by increasing the cutoff, the monopolist forgoes the surplus it would otherwise extract from the marginal type. In a standard monopoly problem, this extractable surplus is $\varphi(\ts)$. In the current setting, however, the foregone surplus is scaled down by the allocation probability at the public option, precisely because a buyer can fall back on the public option if the monopolist chooses to exclude her. 

The right-hand side captures the monopolist's marginal benefit: by increasing the cutoff, the monopolist shifts additional demand to the public option, which lowers its allocation probability and makes the public good a \emph{damaged good}. As a result, guaranteed access from the private market becomes more valuable, which allows the monopolist to extract higher surplus from inframarginal types. Since the public option's allocation probability is also the fraction of virtual surplus that the monopolist cannot extract, an increase in the cutoff that worsens the public option's allocation probability at rate $k/F(\ts)^2$ also raises the extractable share of surplus from each inframarginal type $\theta>\ts$ by the same rate. Using
\[
\int_\ts^{\bar\theta}\varphi(s)dF(s)=\ts(1-F(\ts)),
\]
the right-hand side of \eqref{eq:foc} represents the total marginal benefit from the additional surplus extracted from inframarginal buyers when the public option becomes less attractive.

For each $\theta\in\Theta$, define the function 
\begin{align*}
\omega(\theta)\coloneqq \frac{\varphi(\theta)F(\theta)}{\varphi(\theta)F(\theta)+\theta\big(1-F(\theta)\big)},
\end{align*}
which is continuous and strictly increasing over $[\theta^M, \bar\theta]$ with $\omega(\theta^M)=0$ and $\omega(\bar\theta)=1$. For the subsequent analysis, it will at times be more convenient to rewrite \eqref{eq:foc} so that $\ts$ is the unique solution to 
\begin{equation}
\label{eq:foc_omega}
    \frac{k}{F(\theta)}= \omega(\theta).
\end{equation} 

To gain some intuition for what $\omega(\theta)$ represents, suppose the monopolist marginally increases the cutoff $\theta$. Relative to the standard monopoly problem, excluding the marginal buyer is less costly because the foregone surplus is discounted from $\varphi(\theta)$ to $(1-k/F(\theta))\varphi(\theta)$, as captured by the left-hand side of \eqref{eq:foc}. Thus, $(k/F(\theta))\varphi(\theta)$ is the monopolist's gain from the marginal cost discount. At the same time, raising the cutoff increases the surplus the monopolist extracts from inframarginal buyers, as captured by the right-hand side of \eqref{eq:foc}. Hence, we can interpret $\omega(\theta)$ as the fraction of the total strategic gain from raising the cutoff that comes from the marginal-cost discount channel rather than from the additional rent extraction from inframarginal buyers. This interpretation is useful for the comparative statics in the following section, where the elasticity of $\omega$ governs whether a capacity expansion improves consumer welfare.

\section{Main Results}\label{sec:main}

This section examines two sets of comparative statics. First, I analyze how changes in the capacity of a public option affect consumer welfare in a mixed market. Second, I consider how consumer welfare changes when a monopolist enters a market that was previously served only by a public option.

\subsection{Expanding a Public Option.}\label{sec:comparative}
What is the effect of introducing a public option to a monopoly-only market, or expanding the capacity of an existing public option in a mixed market? For example, in the United States, where the housing market is primarily supplied by the private sector, how would increasing the public housing supply affect consumer welfare?

To address these questions, I examine how each consumer type's payoff varies with the public option's capacity. This granular approach is important because a policymaker may wish to evaluate the impact of the public option from a non-utilitarian perspective, such as a Rawlsian welfare criterion. Alternatively, the policymaker may adopt a utilitarian perspective but use non-uniform welfare weights, such as placing a greater weight on the low-value consumers who rely on the public option and are typically the policy's intended beneficiaries. 

For each $k \in (0, 1)$, let $\ts(k)\in\Theta$ denote the equilibrium cutoff solving \eqref{eq:foc}. Define $\beta(k)\coloneqq k/F\big(\ts(k)\big)$ as the public option's equilibrium allocation probability, and $p^*(k)\coloneqq\ts(k)\left(1-\beta(k)\right)$ as the profit-maximizing monopoly price. The equilibrium payoff of a type-$\theta$ buyer is given by 
\begin{equation*}
U(\theta, k)=
    \begin{cases}
 \theta\,\beta(k)    & \text{if }  \theta<\ts(k)\\
\theta-p^*(k)     & \text{if }  \theta\geq \ts(k)
    \end{cases}.
\end{equation*}

The first result in this section establishes that as the public option's capacity expands, the monopolist's supply (and the set of consumers served by the monopolist) continuously shrinks.

\begin{proposition}
\label{prop:cutoffcompstat}
The cutoff $\ts(k)$ is continuous and strictly increasing in $k$. Furthermore, $\lim_{k\to 0} \ts(k)=\theta^M$ and $\lim_{k\to 1}\ts(k)=\bar \theta$.
\end{proposition}

Because the equilibrium cutoff is continuous in $k$, the allocation probability, price, and payoff functions also vary continuously with $k$. Furthermore, as $k\to 1$, a growing mass of consumers obtain the good from the public option with diminishing congestion. Hence, when the capacity of the public option is sufficiently large, consumers are strictly better off than in a monopoly-only market (i.e., when $k=0$).    
 
At the same time, as $k$ increases, the monopolist serves an increasingly narrower segment of the market. In particular, expanding the capacity of the public option always erodes the monopolist's profits.\footnote{One can easily establish that the monopolist's profit is strictly decreasing in $k$ by applying the envelope theorem to \eqref{eq:profit2}.}

What about consumer welfare? First, consider low-value consumers who rely on the public option. These consumers benefit from a capacity expansion if and only if the public option's allocation probability increases. Here, an expansion generates two opposing effects. First, holding fixed the set of consumers who buy from the monopolist, the additional capacity alleviates rationing at the public option. Second, as established in \autoref{prop:cutoffcompstat}, the monopolist responds to a capacity expansion by raising the cutoff, which shifts even more demand to the public option.

The following proposition establishes that the first effect always dominates, and thus, low-value consumers unambiguously benefit from an increase in the public option's capacity despite the monopolist's strategic response. Moreover, the marginal buyer is also strictly better off after a capacity expansion.

 \begin{proposition}
\label{prop:lowvaluecompstat}
 For all $k', k''\in (0,1)$ with $k'<k''$ and for all $\theta\leq \ts(k')$,
 \[
 U(\theta,k'')\geq U(\theta,k'),
 \]
 with a strict inequality if $\theta>0$. Moreover, $U(\ts(k''), k'')>U(\ts(k'),k')$.
 \end{proposition}

Next, consider high-value consumers who purchase from the monopolist. These consumers benefit from a capacity expansion if and only if it induces the monopolist to lower its price. An expansion generates two opposing effects: First, as implied by \autoref{prop:lowvaluecompstat}, a capacity expansion improves the public option's allocation probability. This makes the public option more attractive for all buyer types, exerting downward pressure on the monopolist's price. Second, as established by \autoref{prop:cutoffcompstat}, a capacity expansion leads to a cutoff increase. Consequently, the marginal buyer has a higher type and thus a higher willingness-to-pay for guaranteed allocation from the monopolist, which creates upward pressure on the price. The following proposition establishes when the downward pressure on prices  dominates.

\begin{proposition}
\label{prop:highvaluecscompstat}
For all $k',k''\in(0,1)$ with $k'<k''$ and all $\theta>\ts(k'')$, 
\[
U(\theta,k'')\geq U(\theta,k')
\]
if and only if $\theta(1-\omega(\theta))$ is non-increasing in $\theta$ on $(\theta^M,\bar\theta)$.
\end{proposition}

Given a cutoff type $\theta$, recall that $1-\omega(\theta)$ represents the share of the monopolist's strategic gain from raising the cutoff that comes from extracting additional surplus from inframarginal types. \autoref{prop:highvaluecscompstat} shows that a capacity expansion lowers the monopoly price, and hence benefits the buyers who remain in the private market, exactly when this share of the strategic gain from raising the cutoff declines sufficiently quickly to offset the increase in the marginal buyer's valuation. 

When $\varphi$ is differentiable, I show (in the proof of the proposition) that the monotonicity condition in the proposition is mathematically equivalent to
\begin{equation}
    \label{eq:condition}
      \frac{\theta\, f(\theta)}{1-F(\theta)}+ \omega(\theta)\cdot \frac{\theta\,  \varphi'(\theta)}{\varphi(\theta)}\geq 2
    \end{equation}
for all $\theta\in(\theta^M, \bar\theta)$. In other words, the price decrease following a capacity expansion arises from a combination of two forces: the loss of market share and the cost of excluding the marginal buyer.

The first force is the sensitivity of the monopolist's demand, $1-F(\theta)$, to changes in the cutoff type, captured by the first term on the left-hand side of \eqref{eq:condition}. A high elasticity of demand implies that the monopolist cannot raise the cutoff without a substantial loss in the mass of inframarginal types from which it extracts surplus.

The second force is how quickly the cost of excluding the marginal buyer rises with the cutoff, captured by the weighted elasticity of the virtual value function in the second term. A highly elastic virtual value function implies that the cost of excluding the marginal type increases rapidly as the cutoff increases. This sensitivity is weighted by $\omega$, which measures how much of the monopolist's benefit comes from the cost discount it enjoys at the marginal type.

Thus, a capacity expansion lowers the monopoly price when either the private market demand thins out quickly enough as the firm raises the cutoff, or the cost of excluding successively higher types rises quickly enough, or both. In these cases, the downward pressure on price arising from the improved public option dominates the upward pressure from the increased valuation of the marginal buyer, thereby benefiting all high-value consumers.

Notice that \autoref{prop:lowvaluecompstat} considers buyers who rely on the public option both before and after the capacity expansion, while \autoref{prop:highvaluecscompstat} considers buyers who purchase from the monopolist both before and after the expansion. A capacity increase from $k'$ to $k''$, however, also changes the market participation of buyer types $\theta\in(\ts(k'),\ts(k''))$ who purchase from the monopolist before the expansion but rely on the public option after the expansion. The following corollary shows that these switching types benefit from a capacity expansion exactly under the same monotonicity condition characterized in \autoref{prop:highvaluecscompstat}.

\begin{corollary}\label{cor:1}
\label{cor:marginalrankings}
For all $k',k''\in(0,1)$ with $k'<k''$ and all $\theta\in[\ts(k'), \ts(k'')]$, 
\[
U(\theta,k'')\geq U(\theta, k')
\]
if and only if $\theta(1-\omega(\theta))$ is non-increasing in $\theta$ on $(\theta^M, \bar\theta)$.
\end{corollary}

\autoref{prop:lowvaluecompstat}, \autoref{prop:highvaluecscompstat}, and \autoref{cor:1} together provide conditions under which an increase in the capacity of the public option improves the payoff of \textit{all} buyer types. Consequently, any aggregation of buyers' payoffs---including the standard utilitarian consumer welfare $C(k)\coloneqq \mathbb{E}_F[U(\theta,k)]$---is increasing in $k$. More broadly, a capacity expansion is especially likely to improve aggregate consumer welfare when the monopoly excludes a sufficiently large fraction of buyers. In such cases, even when Condition \eqref{eq:condition} fails, the welfare gains accruing to low-value consumers offset any losses experienced by high-value consumers, provided the aggregation method does not place disproportionate weight on high-value consumers.

I conclude this section by considering a family of parameterized distributions. Let $F_{a,b}$ be a piecewise-uniform distribution on the unit interval with density
\[
f_{a,b}(\theta)=
\begin{cases}
    b & \text{if } \theta<a,\\[6pt]
\displaystyle\frac{1-ab}{1-a} & \text{if } \theta\geq a.
\end{cases}
\]
When $a\leq 1/2$ and $b\in(0,1]$, the associated virtual value function is monotone and $\theta^M=1/2$.\footnote{Technically, $F_{a,b}$ is not continuously differentiable on $[0,1]$ when $a>0$ and $b<1$. However, one can smooth the jump in the density over a small neighborhood without changing the qualitative conclusions.} Hence, in the monopoly-only benchmark, a mass $F_{a,b}(\theta^M)>0$ of buyers is excluded from the market. Introducing or expanding a public option benefits these low-value buyers, as shown by \autoref{prop:lowvaluecompstat}. For high-value buyers, however, \autoref{prop:highvaluecscompstat} implies that every capacity expansion is beneficial if and only if
\[
\frac{1-ab}{1-a}\leq \frac{4}{3},
\]
where the above inequality comes from \eqref{eq:condition}. 

When $a=0$, $F_{a,b}$ is simply the uniform distribution, and the above inequality is satisfied. In this case, we can derive the equilibrium objects in closed form: the optimal cutoff is $\ts(k)=(1+k)/2$, the allocation probability is $\beta(k)=2k/(1+k)$, and the profit-maximizing price is $p^*(k)=(1-k)/2$. Since $\beta(k)$ rises and $p^*(k)$ falls with $k$, both low- and high-value buyers benefit from a capacity expansion. The standard utilitarian consumer welfare,
\[
C(k)=\frac{1+4k-k^2}{8},
\]
is strictly increasing in $k$. Social welfare, defined as $S(k)\coloneqq C(k)+p^*(k)\big(1-F(\ts(k))\big)$, in this case is given by
\[
S(k)=\frac{3+k^2}{8},
\]
which is also increasing in $k$. This measure, however, abstracts from the cost of financing or providing the public option. A full public-finance evaluation would additionally require specifying the cost of public provision.

Next, consider the case where $a=1/2$ and $b=1/10$. Then $(1-ab)/(1-a)=1.9>4/3$, so high-value buyers do not always benefit from a capacity expansion. Although closed-form solutions are not available, it is straightforward to numerically compute $C(k)$ and show that it is non-monotone: it is strictly decreasing until $k\approx 0.0124$ and strictly increasing thereafter, as shown by the blue curve in \autoref{fig:fail}. Moreover, the monopoly-only benchmark inefficiently excludes types $\theta<\theta^M=1/2$, and introducing a small public option further lowers social welfare. In fact, social welfare is strictly decreasing until $k\approx 0.237$ and strictly increasing thereafter, converging to the efficient allocation as $k\to 1$. Hence, social welfare is also non-monotone in the public option's capacity, as shown by the red curve in \autoref{fig:fail}.

\begin{figure}[t]
\centering
\begin{tikzpicture}[spy using outlines={rectangle, magnification=3.5,  connect spies}]


\pgfmathdeclarefunction{Fdist}{1}{%
    \pgfmathparse{1.9*(#1)-0.9}%
}
\pgfmathdeclarefunction{vvalue}{1}{%
    \pgfmathparse{2*(#1)-1}%
}
\pgfmathdeclarefunction{omegaex}{1}{%
    \pgfmathparse{(vvalue(#1)*Fdist(#1))/(vvalue(#1)*Fdist(#1)+(#1)*(1-Fdist(#1)))}%
}
\pgfmathdeclarefunction{kofcutoff}{1}{%
    \pgfmathparse{Fdist(#1)*omegaex(#1)}%
}
\pgfmathdeclarefunction{Ilow}{1}{%
    \pgfmathparse{0.95*(#1)^2-0.225}%
}
\pgfmathdeclarefunction{Ihigh}{1}{%
    \pgfmathparse{0.725-Ilow(#1)}%
}
\pgfmathdeclarefunction{priceex}{1}{%
    \pgfmathparse{(#1)*(1-omegaex(#1))}%
}
\pgfmathdeclarefunction{Cwelfare}{1}{%
    \pgfmathparse{omegaex(#1)*Ilow(#1)+Ihigh(#1)-priceex(#1)*(1-Fdist(#1))}%
}
\pgfmathdeclarefunction{Swelfare}{1}{%
    \pgfmathparse{omegaex(#1)*Ilow(#1)+Ihigh(#1)}%
}

\begin{axis}[
    name=main,
    width=0.72\textwidth,
    height=0.48\textwidth,
    xmin=0, xmax=1,
    ymin=0.18, ymax=0.75,
    xlabel={Public option capacity $(k)$},
    ylabel={Welfare},
    grid=none,
    grid style={dashed, gray!30},
    legend style={at={(1.2,0.207)}},
    legend cell align=left,
    tick label style={font=\small},
    label style={font=\small},
    title style={font=\small},
]

\addplot[
    blue,
    very thick,
    domain=0.5:0.999999,
    samples=350,
    smooth,
]
({kofcutoff(x)}, {Cwelfare(x)});
\addlegendentry{$C(k)$}

\addplot[
    red,
    very thick,
    domain=0.5:0.999999,
    samples=350,
    smooth,
]
({kofcutoff(x)}, {Swelfare(x)});
\addlegendentry{$S(k)$}

        \coordinate (zoom_point) at (axis cs:0.04, 0.23);
        \coordinate (magnification_pos) at (axis cs:0.72, 0.5);
            \spy [width=3.5cm, height=3cm] on (zoom_point) in node [fill=none, anchor=north] at (magnification_pos);
\end{axis}

\end{tikzpicture}
\captionsetup{oneside,margin={0cm,0cm},justification=justified, singlelinecheck=false}
\caption{Consumer and social welfare as functions of public option capacity for the piecewise-uniform distribution $F_{a,b}$ with $a=1/2$ and $b=1/10$.}
\label{fig:fail}
\end{figure}

\subsection{Introducing a Monopolist.}\label{sec:introduce} What is the effect of introducing a monopolist into a market served exclusively by a congested public option? This question is particularly relevant for markets like the UK’s National Health Service, where public provision is widespread but allocation is rationed due to capacity constraints.

Prior to introducing the monopolist, all buyers rely on the public option. Hence, the allocation probability is exactly $k$, and a type-$\theta$ buyer's payoff is $\theta\,k$. The following proposition establishes that introducing a monopolist to such a market improves the payoff of each consumer type.

\begin{proposition}
\label{prop:highcs}
For all $k\in (0,1)$ and all $\theta\in\Theta$,  $U(\theta, k)\geq \theta\,k$, with a strict inequality if $\theta>0$.
\end{proposition}

The intuition behind \autoref{prop:highcs} is straightforward. High-value consumers benefit from access to the private market, which offers guaranteed allocation rather than a rationed public option. Moreover, as demand shifts from the public option to the private market, the public option’s allocation probability rises. This improves the payoff of low-value consumers who continue to rely on the public option. Therefore, all types benefit from the introduction of the monopolist.

\section{Extensions}\label{sec:extensions}
\subsection{Oligopolistic Mixed Markets.}\label{sec:oligopoly}
In the baseline model, the private market is served by a single firm. However, mixed markets often feature multiple private providers competing against a public option. To examine how competition alters the strategic incentives to congest the public option, suppose the private market consists of $N \ge 1$ identical, zero-marginal-cost firms competing in quantities (Cournot competition).

Let $q_i \ge 0$ denote the quantity supplied by firm $i \in \{1, \dots, N\}$, let $Q = \sum_{i=1}^N q_i$ denote the total private market supply, and let $Q_{-i}=\sum_{j\neq i}q_j$ denote the total private market supply absent firm $i$. Because the private market clears by serving the highest-valuation buyers, any total quantity $Q$ determines a marginal buyer type $\tc(Q)$ who purchases at price $P(Q)$, where $1-F(\tc(Q))=Q$ and $P(Q)$ is given by \eqref{eq:market clearing price}. Each firm $i$, taking $Q_{-i}$ as given, chooses $q_i$ to maximize its profit:
\[
\max_{q_i\geq 0} P(q_i+Q_{-i})\, q_i.
\]

As a first step toward characterizing the Cournot equilibrium outcomes, note that a continuum of zero-profit equilibria exists when $N\geq 2$. This is because the price $P(Q)$ that clears the private market is equal to zero whenever $Q \geq 1-k$. Consequently, if $Q_{-i} \geq 1-k$, the market price remains at $P=0$ regardless of firm $i$'s supply. Thus, for $N\geq 2$, any strategy profile where each firm $i$ produces $q_i \geq (1-k)/(N-1)$ constitutes a Nash equilibrium. Importantly, this continuum of zero-profit equilibria arises only because firms have zero marginal cost. If firms face even an arbitrarily small positive marginal cost of production, these equilibria vanish. 

Henceforth, I focus on equilibria in which firms earn positive profit. Moreover, I restrict attention to symmetric equilibria where $q_i=q$ for all $i\in \{1,\ldots, N\}$. 

Recall that $\theta^M$ is the marginal buyer type in a standard monopoly market,  and $\ts$ (from \autoref{prop:optimal}) is the marginal buyer type when the monopoly competes with the public option. Let 
\[
\varphi_N(\theta)\coloneqq \frac{1}{N}\,\varphi(\theta)+\frac{N-1}{N}\,\theta,
\]
be the \emph{competition-adjusted virtual value function}. The weight $1/N$ corresponds to the standard Cournot \textit{conduct parameter} and captures a firm's market power. Observe that $\varphi_N$ is strictly increasing and  $\varphi_N(\theta)\geq \varphi(\theta)$ for all $\theta$. In a standard $N$-firm oligopoly market, the symmetric Cournot equilibrium marginal buyer type is given by
\[
\theta^O_N\coloneqq \min\{\theta\in\Theta:\varphi_N(\theta)\geq 0\},
\]
with each firm supplying $q=(1-F(\theta^O_N))/N$. Finally, observe that $\theta^O_N\leq\theta^M$ for all $N\geq 1$ with equality holding when $N=1$. 

\begin{proposition}\label{prop:cournot}
There exists a unique symmetric equilibrium of the $N$-firm Cournot oligopoly in which the firms earn positive profits. This equilibrium is characterized by a cutoff type $\theta^*_N \in \Theta$ that uniquely solves 
\begin{equation}
\label{eq:cournot_foc}
\left(1-\frac{k}{F(\theta)}\right)\varphi_N(\theta)=\frac{1}{N}\frac{k}{F(\theta)^2}\,\theta\big(1-F(\theta)\big).
\end{equation}
In this equilibrium, the cutoff type satisfies $\max\{\theta^O_N, F^{-1}(k)\}<\theta^*_N< \ts$, and
\begin{enumerate}[$(\alph*)$]
    \item Each firm supplies a quantity $q^*=(1-F(\theta^*_N))/N$ at price $P(1-F(\theta^*_N))$,
    \item Buyer types $\theta\geq \theta^*_N$ purchase from the private market, and 
    \item Buyer types $\theta< \theta^*_N$ rely on the public option.
\end{enumerate}
Furthermore, the cutoff type $\theta^*_N$ is strictly decreasing in $N$, with $\lim_{N \to \infty} \theta^*_N = F^{-1}(k)$.
\end{proposition}

\autoref{prop:cournot} demonstrates that the qualitative insights of the monopoly model persist under oligopoly. Consumers are once again segmented into high-value types ($\theta>\theta^*_N$) who purchase from the private market and low-value types ($\theta<\theta^*_N$) who rely on the public option. 

The equilibrium cutoff $\theta^*_N$ uniquely solves \eqref{eq:cournot_foc}, which captures each firm $i$'s tradeoff from a marginal increase in the cutoff type, taking the symmetric strategies of the other firms as given. Relative to \eqref{eq:foc} in the monopoly case, the marginal cost of increasing the cutoff---the left-hand side of \eqref{eq:cournot_foc}---is higher, while the marginal benefit of increasing the cutoff---the right-hand side of \eqref{eq:cournot_foc}---is lower. Intuitively, raising the cutoff is more costly for an individual Cournot firm than a monopolist, since each  firm only partially controls the aggregate private market supply. At the same time, the benefit from worsening the public option is shared equally across all $N$ firms. Consequently, an individual firm's incentive to withhold supply is attenuated relative to the monopoly case.

Formally, for all $N\geq 2$, the equilibrium cutoff $\theta^*_N$ is strictly lower than the monopoly cutoff $\ts$. Hence, the aggregate supply of the private market under Cournot competition is higher than it would be under a monopoly. Yet, it is still the case that $F(\theta^*_N)>k$ for all finite $N$, guaranteeing that the public option remains rationed in any oligopolistic market. 

As the number of firms $N$ increases, the private market expands its aggregate supply, driving down the equilibrium cutoff. Thus, for a fixed capacity $k$, an increase in $N$ raises the equilibrium allocation probability $k/F(\theta^*_N)$ and lowers the equilibrium price $P(1-F(\theta^*_N))$, which strictly improves the welfare of both low- and high-value consumers. As the private market converges to a perfectly competitive limit ($N \to \infty$), the strategic incentive to restrict supply vanishes entirely. The private market expands until the price drops to zero, at which point $\theta^*_\infty = F^{-1}(k)$, meaning the public option clears exactly at capacity and therefore without rationing.

Moreover, the comparative statics regarding the public option's capacity expansion in a mixed market with a monopoly supplier extend naturally to a mixed market with an arbitrary number of private suppliers. For each $N\geq 1$ and $k \in (0,1)$, let $\theta^*_N(k)\in\Theta$ denote the equilibrium cutoff solving \eqref{eq:cournot_foc}. Define $\beta_N(k)\coloneqq k/F\big(\theta^*_N(k)\big)$ as the public option's equilibrium allocation probability, and $p^*_N(k)\coloneqq\theta^*_N(k)\left(1-\beta_N(k)\right)$ as the equilibrium price. The equilibrium payoff of a type-$\theta$ buyer is given by 
\[ 
U_N(\theta, k)=  
\begin{cases}
  \theta\,\beta_N(k)  & \text{if }  \theta<\theta^*_N(k)\\
  \theta-p^*_N(k)  & \text{if }  \theta\geq \theta^*_N(k)
\end{cases}.
\]

The following proposition generalizes the comparative statics of a monopoly mixed market (\autoref{prop:cutoffcompstat}-\autoref{prop:highvaluecscompstat},  as well as \autoref{cor:1}) to any number $N\geq 1$ of private suppliers. To state the result, define the \emph{competition-adjusted weight function} by
\[
\omega_N(\theta)\coloneqq\frac{\varphi_N(\theta)F(\theta)}{\varphi_N(\theta)F(\theta)+\frac{1}{N}\theta\big(1-F(\theta)\big)}.
\]
For $N=1$, observe that $\omega_N(\theta)=\omega(\theta)$ for all $\theta\in\Theta$, where $\omega$ is as defined in \eqref{eq:foc_omega}. Moreover, for all $N\geq 1$, $\omega_N(\theta)$ is continuous and strictly increasing over $(\theta^O_N, \bar\theta)$ with $\omega_N(\theta^O_N)=0$ and $\omega_N(\bar\theta)=1$. Finally, for each $N\geq 1$ and $k\in (0,1)$, observe that the equilibrium cutoff $\theta^*_N(k)$ is the unique solution to 
\begin{equation}
\label{eq:foc_omega_cournot}
    \frac{k}{F(\theta)}= \omega_N(\theta),
\end{equation}
which we derive by rearranging \eqref{eq:cournot_foc}.

\begin{proposition}
    \label{prop:cournot_comparative}
For any $N\geq 1$, the following are true:
\begin{enumerate}[$(\alph*)$]
    \item The equilibrium cutoff $\theta^*_N(k)$ is continuous and strictly increasing in $k$, with $\lim_{k\to 0}\theta^*_N(k)=\theta^O_N$ and $\lim_{k\to 1}\theta^*_N(k)=\bar\theta$. 
    \item  For all $k', k''\in (0,1)$ with $k'<k''$ and all $\theta< \theta^*_N(k')$,  \[
    U_N(\theta,k'')\geq U_N(\theta,k')
    \]
    with a strict inequality if $\theta>0$. Furthermore, $U_N(\theta^*_N(k''), k'')>U_N(\theta^*_N(k'),k')$.
    \item For all $k', k''\in (0,1)$ with $k'<k''$ and all $\theta\geq\theta^*_N(k')$, 
    \[
    U_N(\theta, k'')\geq U_N(\theta, k')
    \]
    if and only if $\theta(1-\omega_N(\theta))$ is non-increasing in $\theta$ on $(\theta^O_N, \bar \theta)$.
\end{enumerate}
\end{proposition}

\subsection{Private Market as a Complement to Public Option.}\label{sec:complement}
In the baseline model, the monopoly competes with the public option. This is formalized in the timing of the baseline game in which buyers are forced to choose between the public option and the monopolist. In this section, I consider a setting in which buyers turn to the monopolist only if they were not lucky enough to receive the good for free from the public option. Thus, the private market here serves as a complement to the public option, with buyers using it to ``top up'' any demand unmet by the public option.

Formally, I consider the following alternate timing: First, the public option's capacity $k$ is established, and each buyer privately observes her type. Second, all buyers approach the public option, which allocates the good for free, using a lottery whenever demand exceeds capacity. Third, taking the remaining buyers as given, the monopolist posts a price $p$ to maximize its profit. Finally, any buyer who did not receive a good from the public option chooses whether to purchase from the monopoly at the posted price.

Notice that in this setting, the monopolist is unable to affect the public option's demand through its own strategic pricing. Since all buyers first approach the public option and there is a unit mass of consumers, the allocation probability at the public option is exactly $k$. This implies that a mass $1-k$ of buyers remain in the market after the public option has been allocated. Furthermore, because the public option is allocated at random across all buyers, the type distribution of buyers who remain in the market is $F$. 

Consequently, the monopolist faces a standard monopoly problem, making the optimal posted price $\theta^M$, regardless of the public option's capacity. In this case, low-value types ($\theta<\theta^M$) rely only on the public option while high-value types ($\theta> \theta^M$) top up their demand in the private market. Hence, the equilibrium payoff of a type-$\theta$ consumer is given by 
\[
\widetilde{U}(\theta, k)=\begin{cases}
    \theta\, k  & \text{if } \theta<\theta^M\\
\theta\,k+(\theta-\theta^M)(1-k)     & \text{if } \theta\geq \theta^M
\end{cases}.
\]

Let us first consider how expanding the public option's capacity affects consumers in this setting. As is clear from the above expression, the payoff of low-value buyers ($\theta< \theta^M$) is increasing in $k$. This is similar to the comparative statics established in \autoref{prop:lowvaluecompstat}. Moreover, in contrast to \autoref{prop:highvaluecscompstat}, a capacity expansion in this setting also increases the payoff of high-value buyers ($\theta\geq  \theta^M$) without additional conditions on the market primitives. 

Intuitively, since the monopolist can no longer induce varying demand at the public option, expanding the public option's capacity improves its allocation probability without giving rise to any offsetting strategic behavior from the monopolist. Therefore, all buyers are more likely to have their demand met through a free public option, reducing their need to top up demand in the private market. 

Next, consider the impact of introducing a complementary monopoly to a market initially served by a standalone public option. Recall from \Cref{sec:introduce} that type-$\theta$'s payoff in a market served only by a public option equals $\theta\,k$. Hence, the introduction of a complementary monopoly is strictly beneficial only to high-value buyers, which stands in contrast to \autoref{prop:highcs} in which (almost) all buyer types strictly benefit. 

Finally, consider buyers’ payoffs under the baseline regime and the complement regime. Which regime is better for buyers: one in which the private market competes with the public option, or one in which it complements it? This question is relevant if, for example, the government (or  planner) controls not just the supply of the public option but also the timing of its allocation. 

To answer this question, fix an arbitrary public option capacity $k\in(0,1)$. Since the same buyer type may be considered a high-value type under one regime but a low-value one in another, I consider three distinct categories of buyers: \textit{always-low-value} buyers ($\theta\leq \theta^M$) who never purchase from the monopolist; \textit{always-high-value} buyers ($\theta\geq \ts(k)$) who are willing to purchase from the monopolist in both regimes; and \textit{switchers} ($\theta^M<\theta<\ts(k)$) who are willing to purchase from the monopolist only in the complement regime.\footnote{Since the optimal cutoff in the baseline model $\ts(k)$ is strictly increasing in $k$ with $\ts(0)=\theta^M$ (\autoref{prop:cutoffcompstat}), there are no types that purchase from the monopolist only under the baseline regime.}

\begin{figure}[htbp]
\centering
\begin{tikzpicture}[scale=1.1]

    \def\lowbound{0}
    \def\midsplit{3.5}
    \def\highsplit{6}
    \def\upbound{10}

    \draw[ultra thick] (\lowbound, .75) -- (\upbound, .75);
    \node[left] at (-0.5, .75) {\textbf{Baseline:}};
    
    \draw (\lowbound, .6) -- (\lowbound, .9) node[above] {$\underline{\theta}$};
    \draw (\highsplit, .6) -- (\highsplit, .9) node[above] {$\ts(k)$};
    \draw (\upbound, .6) -- (\upbound, .9) node[above] {$\bar{\theta}$};
    
    \node[above=1pt] at ({(\lowbound+\highsplit)/2}, .75) {\small Public Option};
    \node[above=1pt] at ({(\highsplit+\upbound)/2}, .75) {\small Private Market};

    \draw[ultra thick] (\lowbound, 0) -- (\upbound, 0);
    \node[left] at (-0.5, 0) {\textbf{Complement:}};
    
    \draw (\lowbound, 0.15) -- (\lowbound, -0.15) node[below] {$\underline{\theta}$};
    \draw (\midsplit, 0.15) -- (\midsplit, -0.15) node[below] {$\theta^M$};
    \draw (\upbound, 0.15) -- (\upbound, -0.15) node[below] {$\bar{\theta}$};

    \node[below=1pt] at ({(\lowbound+\midsplit)/2}, 0) {\small Public Option};
    \node[below=1pt] at ({(\midsplit+\upbound)/2}, 0) {\small Public Option + Private Market};

        \draw[ultra thick, red] (\lowbound, .75) -- (\highsplit, .75);
        \draw[ultra thick, blue] (\highsplit, .75) -- (\upbound, .75);

        \draw[ultra thick, red] (\lowbound, 0) -- (\midsplit, 0);
        \draw[ultra thick, blue] (\midsplit, 0) -- (\upbound, 0);

        \draw[dashed, gray, thick] (\midsplit, 0) -- (\midsplit, .75);
         \draw[dashed, gray, thick] (\highsplit, 0) -- (\highsplit, .75);
    \begin{scope}[decoration={brace, mirror, amplitude=7pt}]
        
        \draw[decorate] (\lowbound, -0.8) -- (\midsplit, -0.8) 
            node[midway, below=8pt, font=\small, align=center] {Always-low-value};
            
        \draw[decorate] (\midsplit, -0.8) -- (\highsplit, -0.8) 
            node[midway, below=8pt, font=\small, align=center] {Switchers};
            
        \draw[decorate] (\highsplit, -0.8) -- (\upbound, -0.8) 
            node[midway, below=8pt, font=\small, align=center] {Always-high-value};
            
    \end{scope}

\end{tikzpicture}
\caption{Classification of buyer types across regimes.}
\end{figure}

At first glance, one might expect buyers to be better off under the complement regime, which does not force them to choose between the public option and the private market. However, for any type $\theta\leq\theta^M$, a direct comparison of payoffs under the baseline regime $U(\theta, k)$ and the complement regime $\widetilde U(\theta, k)$ reveals that
\[
U(\theta,k)=\theta\,\beta(k)>\theta\, k=\widetilde U(\theta, k),
\]
where the inequality follows from the fact that $\ts(k)<\bar\theta$ (\autoref{prop:optimal}) and $\beta(k)=k/F(\ts(k))$. Thus, always-low-value buyers are worse off under the complement regime than under the baseline regime. These types rely exclusively on the public option in either regime, so the additional congestion generated by the always-high-value buyers under the complement regime strictly harms them. Consequently, under the complement regime, the very types that the public option is typically designed to serve are crowded out by higher-type buyers.

Moreover, since both $U(\theta,k)$ and $\widetilde U(\theta, k)$ are continuous in $\theta$, and since $U(\theta^M,k)>\widetilde U(\theta^M, k)$, there always exists a set of switchers who are also worse off under the complement regime than under the baseline regime. More broadly, whether all types are worse off under the complement regime depends on the type distribution, as formalized by the following necessary and sufficient condition.

\begin{proposition}\label{prop:complement_vs_baseline}
Suppose that for all $\theta\in(\theta^M,\bar \theta)$, 
\begin{equation}\label{eq:complement condition}
    \omega(\theta)\geq \frac{\theta-\theta^M}{\theta-\theta^M F(\theta)}.
\end{equation}
Then for all $k\in (0,1)$ and all $\theta\in\Theta$, $U(\theta,k)\geq \widetilde U(\theta,k)$. Alternatively, if the above inequality fails for some  $\theta\in (\theta^M, \bar \theta)$, then there exists an open set $K\subseteq (0,1)$ and a cutoff type $\hat\theta(k)\in (\theta^M, \ts(k))$ for each $k\in K$ such that $U(\theta,k)>\widetilde U(\theta,k)$ for all $\theta<\hat\theta(k)$ and $U(\theta,k)< \widetilde U(\theta,k)$ for all $\theta>\hat\theta(k)$.
\end{proposition}

Intuitively, switchers and always-high-value types benefit from the ability to ``double-dip'' under the complement regime. On the other hand, they benefit in the baseline regime from a less congested public option and possibly lower prices due to the competitive pressure exerted by the public option on the private market. Which of these benefits dominates depends on the type distribution.

When Condition \eqref{eq:complement condition} is satisfied, for any $k \in (0,1)$, the baseline-regime price $p^*(k)$, relative to the complement-regime price $\theta^M$, is low enough to more than compensate the always-high-value types for the loss of their private top-up option. Furthermore, this low price draws enough high-value demand away from the public option to raise its allocation probability sufficiently, which in turn compensates the switchers for the loss of their top-up option. 

The comparison between the baseline and the complement regimes highlights the public option's role as a screening device. In the baseline regime, buyers must choose between the public option and the private market. This induces high-valuation buyers to self-select into the private market, leaving the public option less congested for low-valuation buyers. In the complement regime, by contrast, the public option's role as a screening device is shut down since all consumers first try the public option, leading high-valuation buyers to crowd out the lower-valuation buyers whom the public option is primarily intended to serve. Thus, policy design in mixed markets should account not only for public capacity, but also for how the timing of access to the public option affects which consumer types sort into which markets.

I conclude this subsection with the two examples from \Cref{sec:comparative}. Consider first the case where the type distribution is $F_{a,b}$ with $a=0$, which is simply the uniform distribution over the unit interval. In this case, $\theta^M=1/2$, $\omega(\theta)=(2\theta-1)/\theta$, and 
\[
\frac{\theta-\theta^M}{\theta-\theta^M F(\theta)}=\frac{2\theta-1}{\theta}.
\]
Hence, \eqref{eq:complement condition} is satisfied with equality for all $\theta\in(1/2,1)$, implying that no buyer type prefers the complement regime to the baseline, regardless of the public option capacity. 

Next, consider the piecewise-uniform distribution $F_{a,b}$ with $a=1/2$ and $b=1/10$. In this case, over the relevant interval $\theta\in(1/2,1)$, we have $\theta^M=1/2$,
\[
\omega(\theta)=\frac{(2\theta-1)(19\theta-9)}{19\theta^2-18\theta+9},
\]
and
\[
\frac{\theta-\theta^M}{\theta-\theta^M F(\theta)}=\frac{10(2\theta-1)}{\theta+9}.
\]
Since,
\[
\frac{10}{\theta+9}>\frac{19\theta-9}{19\theta^2-18\theta+9}
\]
for all $\theta\in(0,1)$, Condition \eqref{eq:complement condition} fails for all $\theta\in(\theta^M,\bar\theta)$. It follows from the proof of \autoref{prop:complement_vs_baseline} that for every $k\in(0,1)$, the corresponding always-high-value types and some of the switchers prefer the complement regime over the baseline.

\subsection{Fringe Competitive Markets.}\label{sec:fringe}
The strategic effects in a mixed market have a natural counterpart in settings where a monopolist (or dominant firm) competes against a competitive fringe. In these settings, the monopolist supplies a good at zero cost, while the fringe supplies a good---possibly of inferior quality---at a positive marginal cost. In the most basic iteration of this model, the fringe's marginal cost is constant, resulting in perfectly elastic supply. As a result, the monopolist cannot influence the fringe's market-clearing price by restricting its own supply and shifting more consumers to the competitive market. Consequently, there is no channel through which the monopolist can reduce consumers’ value in the fringe market. Instead, the competitive market's fixed price exerts downward pressure on the monopolist in the form of a price ceiling.

This stands in sharp contrast to the mixed market model, where the monopolist can reduce consumers’ value of the public option by restricting its own supply and thereby inducing rationing. This mechanism bears a closer resemblance to a fringe model in which competitive firms face increasing marginal costs. In this case, the fringe market features an upward-sloping supply curve, so that a reduction in the monopolist's supply shifts demand to the competitive market and drives the fringe's market-clearing price higher. Thus, both in the mixed market and the fringe market with increasing marginal cost, the monopolist can soften competition by manipulating the attractiveness of the alternative supplier.

However, a competitive fringe, even one with increasing marginal costs, always clears excess demand via prices. Consequently, the good is allocated to consumers with the highest willingness to pay. In contrast, the capacity-constrained public option in the baseline model clears excess demand through random rationing, which introduces allocative inefficiencies. Hence, while these two frameworks could generate similar strategic behavior by the monopolist, the conclusions on consumer welfare in the baseline model need not extend to the fringe model.

\section{Conclusion}\label{sec:conclusion}
This paper studies how a private market competes with a free but capacity-constrained public option. In the baseline model, I show that a profit-maximizing monopolist has an incentive to restrict its own supply in order to induce rationing at the public option and increase consumers’ willingness to pay for guaranteed access. The resulting mixed-market equilibrium is characterized by a cutoff: high-valuation buyers purchase from the monopolist, while low-valuation buyers rely on the congested public option.

This mechanism yields several implications for policy. First, introducing a public option into a monopolistic market, or expanding the capacity of an existing public option, need not improve consumer welfare. Although a capacity expansion always benefits the lower-valuation consumers who rely on the public option, it may also induce the monopolist to raise its price, thereby reducing surplus for higher-valuation consumers. As a result, a public-option intervention can lower aggregate consumer welfare unless it is sufficiently large or the type distribution satisfies an elasticity condition under which the monopoly price falls with capacity. Second, introducing a profit-maximizing monopolist into a market served only by a capacity-constrained public option unambiguously improves consumer surplus for all buyer types, because guaranteed private access draws demand away from the public option and eases rationing.

I also show that the main results are robust to changes in market structure. Under oligopoly, firms still have an incentive to induce rationing at the public option, but this incentive weakens with competition and vanishes in the competitive limit. 

The timing of access to the public option also matters. Requiring consumers to choose between public and private provision can induce higher-valuation consumers to sort into the private market, leaving the public option less congested for the low-valuation consumers it is often intended to serve. By contrast, allowing consumers to access the public option first and then top up in a complementary private market weakens this screening role and results in high-valuation consumers crowding out lower-valuation consumers at the public option. Thus, while allowing consumers to access a public option and a complementary private market may appear more generous, consumers may be better off when the private market competes with, rather than complements, the public option. 

More generally, the appendix characterizes the monopolist’s optimal mechanism in a richer environment in which the public option may offer a lower-quality good at a positive price. When the public option remains sufficiently attractive---for example, when it offers a good of comparable quality at a highly discounted price---the baseline insights continue to apply. When it is sufficiently unattractive, the monopolist instead prefers to undercut the public option’s price rather than congest it, leading to few, if any, consumers relying on the public option.

Taken together, these findings highlight that public provision in mixed markets cannot be evaluated in isolation from firms' strategic responses. A public option may serve as a safety net, but when its allocation is capacity-constrained, it can also become an object of strategic manipulation by private suppliers. Thus, policy targeting the public provision of goods and services in mixed markets should account not only for the direct effects of providing free or inexpensive access to consumers, but also for how the capacity and timing of access to public provision shape private-market behavior. The framework developed here hopefully provides a tractable foundation for future work, for example, studying the design of optimal tax schemes to finance the costly provision of public goods in mixed markets.

\begin{appendix}
\section{Appendix}\label{sec:appendix}
 \titleformat{\subsection}[block]{\normalfont\bfseries\filcenter}{\thesubsection.}{.5em}{}
  \titleformat{\subsubsection}[runin]{\normalfont\bfseries\filcenter}{\thesubsubsection.}{.5em}{}

\subsection{Proofs}
\begin{proof}[Proof of \autoref{prop:optimal}] From \eqref{eq:profit2}, it is clear that the monopolist earns positive profit if and only if it chooses a cutoff $\theta\in (F^{-1}(k),\bar\theta)$, so any profit-maximizing cutoff must lie in this interval. Moreover, any optimal cutoff $\theta$ must satisfy the first-order condition associated with \eqref{eq:profit2}:
\begin{align*}
r(\theta)\coloneqq\frac{k}{F(\theta)^2}\left[\theta\,\big(1-F(\theta)\big)+\varphi(\theta)F(\theta)\right]-\varphi(\theta)=0,
\end{align*}
which is equivalent to Condition \eqref{eq:foc}. 

Notice that $r(\cdot)$ is continuous on $(\underline\theta, \bar\theta]$. Consider first any $\theta\leq F^{-1}(k)$. We then have
\begin{align*}
  r(\theta)&\geq \frac{1}{F(\theta)}\left[\theta\,\big(1-F(\theta)\big)+\varphi(\theta)F(\theta)\right]-\varphi(\theta)=\frac{\theta \big(1-F(\theta)\big)}{F(\theta)}>0,
\end{align*}
where the first inequality follows because $F(\theta)\leq k$ by assumption. Thus, there is no solution to \eqref{eq:foc} on the interval $[\underline\theta, F^{-1}(k)]$.

Next, consider any $F^{-1}(k)<x<y<\bar\theta$. Using the fact that 
\[
\theta\,\big(1-F(\theta)\big)=\int^{\bar\theta}_\theta \varphi(s)dF(s),
\]
we have
\begin{align*}
r(y)-r(x)\leq &\frac{k}{F(y)^2}\left[-\int^{y}_{x}\varphi(s)dF(s)+\varphi(y)F(y)-\varphi(x)F(x)\right]-\varphi(y)+\varphi(x)\\[6pt]
=&\frac{k}{F(y)^2}\int^{y}_{x}F(s)d\varphi(s) -\varphi(y)+\varphi(x)\\[6pt]
<&\frac{k}{F(y)}\int^{y}_{x}d\varphi(s) -\varphi(y)+\varphi(x)\\[6pt]
=&\big(\varphi(y)-\varphi(x)\big)\left(\frac{k}{F(y)}-1\right)\\[6pt]
<&0,
\end{align*}
where the first inequality follows because $F(x)<F(y)$ and because 
\[
\theta\,\big(1-F(\theta)\big)+\varphi(\theta)F(\theta)= \int^{\bar\theta}_{\theta}\varphi(s)dF(s)+\varphi(\theta)F(\theta)>\mathbb{E}_F[\varphi]=\underline \theta\geq 0
\]
for all $\theta>\underline\theta$, the first equality follows from integration by parts, and the second inequality follows because $F(s)/F(y)<1$ for all $s\in (x,y)$ and because $d\varphi(s)\geq 0$. Hence, $r$ is strictly decreasing on $[F^{-1}(k),\bar\theta]$. Moreover,
\[
r(F^{-1}(k))=\frac{F^{-1}(k)(1-k)}{k}>0>\varphi(\bar\theta)(k-1)=r(\bar\theta).
\]
Hence, there is a unique cutoff $\ts\in(F^{-1}(k), \bar\theta)$ that solves \eqref{eq:foc}. We have thus established that $\ts$ is the unique solution to \eqref{eq:foc} over $\Theta$, and consequently, the unique maximizer of \eqref{eq:profit2}.

Finally, let us show $\ts>\theta^M$. This is trivially true if $\theta^M\leq F^{-1}(k)$, so let us focus instead on the case that $\theta^M> F^{-1}(k)$. Since $F^{-1}(k)>\underline \theta$ for $k>0$, this implies that $\theta^M>\underline \theta$. This in turn implies $\varphi(\theta^M)=0$. Since the right-hand side of \eqref{eq:foc} evaluated at $\ts$ is positive, we have $\varphi(\ts)>0$. Hence, by strict monotonicity of the virtual value function, $\ts> \theta^M$.
\end{proof}\bigskip

\noindent\begin{proof}[Proof of \autoref{prop:cutoffcompstat}]
For a given type $\theta\in \Theta$ and capacity $k\in [0,1]$, define
\[
r(\theta,k)\coloneqq \frac{k}{F(\theta)^2}\left[\theta\,\big(1-F(\theta)\big)+\varphi(\theta)F(\theta)\right]-\varphi(\theta),
\]
which is continuous in both arguments. Moreover, for each $k\in (0,1)$, the optimal cutoff type $\ts(k)$ satisfies $r(\ts(k), k)=0$. Since $\ts(k)$ is the unique such solution, we immediately obtain the continuity of the mapping $k\mapsto \ts(k)$ over $(0,1)$.

For each $k\in (0,1)$, the mapping $\theta\mapsto r(\theta, k)$ is strictly decreasing over $(F^{-1}(k), \bar \theta)$, as established in the proof of \autoref{prop:optimal}. Furthermore, for each $\theta\in (\underline \theta, \bar \theta)$, 
\[
\frac{\partial r(\theta, k)}{\partial k}=\frac{1}{F(\theta)^2}\left[\theta\,\big(1-F(\theta)\big)+\varphi(\theta)F(\theta)\right]>0.
\]
Thus, for any $k',k''\in (0,1)$ with $k''>k'$, 
\[
0=r(\ts(k'), k')=r(\ts(k''), k'')>r(\ts(k''), k'),
\]
which implies that $\ts(k'')>\ts(k')$, as desired. 

Let $\ts(1)\coloneqq \lim_{k\to 1}\ts(k)$ and $\ts(0)\coloneqq \lim_{k\to 0}\ts(k)$. To establish the first limit result, note that $\ts(k)\in (F^{-1}(k), \bar \theta)$ for all $k\in (0,1)$. Thus, $\bar \theta=\lim_{k\to 1}F^{-1}(k)\leq\ts(1)\leq \bar \theta$, so we conclude that $\bar\theta=\ts(1)$, as desired.

To establish the second limit result, note that $\ts(k) \geq \theta^M$ for all $k \in (0,1)$, which implies $\ts(0) \geq \theta^M$. Suppose, for the sake of contradiction, that $\ts(0) > \theta^M$. Thus, $\ts(0) > \underline{\theta}$ and $F(\ts(0)) > 0$. As a result, $(\theta,k)\mapsto r(\theta, k)$ is continuous at the point $(\ts(0), 0)$. Since $r(\ts(k), k)=0$ for all $k \in (0,1)$, taking the limit as $k \to 0$, we obtain $r(\ts(0),0)=0$. On the other hand, by definition, $r(\ts(0),0)=-\varphi(\ts(0))<0$, where the inequality follows from the strict monotonicity of $\varphi$ and the assumption that $\theta^M<\ts(0)$, yielding a contradiction. Thus, it must be that $\ts(0)=\theta^M$, as desired. 
\end{proof}\bigskip

\noindent\begin{proof}[Proof of \autoref{prop:lowvaluecompstat}]
From \autoref{prop:cutoffcompstat},  the mapping $k\mapsto \ts(k)$ is continuous and strictly increasing. Given $k\in (0,1)$, recall that for all $\theta\leq \ts(k)$, the buyer's payoff is $U(\theta, k)=\theta\,\beta(k)$. Hence, for both statements of the proposition, it suffices to prove that $\beta(k)$ is a strictly increasing function.

To see this, note that $\beta(k)=\omega(\ts(k))$ from \eqref{eq:foc_omega}. As established, $\omega(\theta)$ is continuous and strictly increasing in $\theta$ over $[\theta^M, \bar\theta]$. Because $\ts(k)$ is strictly increasing in $k$ with $\lim_{k\to 0}\ts(k) = \theta^M$ and $\lim_{k\to 1}\ts(k) = \bar{\theta}$, it immediately follows that the composition $\beta(k) = \omega(\ts(k))$ is also strictly increasing.
\end{proof}\bigskip

\noindent\begin{proof}[Proof of \autoref{prop:highvaluecscompstat}]
Given $k\in (0,1)$, recall that for all $\theta>\ts(k)$, the buyer's payoff is $U(\theta, k)=\theta-p^*(k)$, where $p^*(k)=\ts(k)\big(1-\beta(k)\big)$. Thus, if $p^*(k)$ is non-increasing over $(0,1)$, then for all $k',k''\in (0,1)$ with $k'<k''$ and all $\theta>\ts(k'')$, $U(\theta,k'')\geq U(\theta,k')$. Conversely, if $p^*(k)$ is strictly increasing over some open set $K\subseteq (0,1)$, then for all $k',k''\in K$ with $k'<k''$ and all $\theta>\ts(k'')$, $U(\theta,k'')<U(\theta,k')$.

From \autoref{prop:cutoffcompstat}, the mapping $k\mapsto \ts(k)$ is continuous and strictly increasing. From \eqref{eq:foc_omega}, we can equivalently express the price as $p^*(k)=\ts(k)\big(1-\omega(\ts(k))\big)$. Because $\ts(k)$ is continuous in $k$ with $\lim_{k\to 0}\ts(k) = \theta^M$ and $\lim_{k\to 1}\ts(k) = \bar{\theta}$, the mapping $k\mapsto p^*(k)$ is non-increasing over $(0,1)$ if and only if $\theta\mapsto \theta(1-\omega(\theta))$ is non-increasing over $(\theta^M, \bar\theta)$. 

When $\varphi$ is differentiable, then the monotonicity condition is equivalent to
\[
\frac{\theta\,\omega'(\theta)}{1-\omega(\theta)}\geq 1
\]
for all $\theta\in(\theta^M, \bar\theta)$. Let me conclude the proof by showing that the above inequality is also equivalent to \eqref{eq:condition}. Notice that 
\begin{align*}
    \omega'(\theta)&=\frac{(1-\omega(\theta))\varphi'(\theta) F(\theta)+\varphi(\theta)f(\theta)}{\varphi(\theta)F(\theta)+\theta\,\big(1-F(\theta)\big)}\\\\
    &=\frac{(1-\omega(\theta))\varphi'(\theta) F(\theta)}{\varphi(\theta)F(\theta)+\theta\,\big(1-F(\theta)\big)}\cdot \frac{\varphi(\theta)}{\varphi(\theta)}+\frac{\varphi(\theta)f(\theta)}{\varphi(\theta)F(\theta)+\theta\,\big(1-F(\theta)\big)}\cdot\frac{\theta(1-F(\theta))}{\theta(1-F(\theta))}\\\\
    &=(1-\omega(\theta))\left[\frac{\varphi'(\theta)}{\varphi(\theta)}\cdot \omega(\theta)+\frac{\varphi(\theta)f(\theta)}{\theta(1-F(\theta))}\right],
\end{align*}
where the last line follows from the definition of $\omega$. Hence, 
\begin{align*}
    \frac{\theta\,\omega'(\theta)}{1-\omega(\theta)}&=\frac{\theta\, \varphi'(\theta)}{\varphi(\theta)}\cdot \omega(\theta)+\frac{\varphi(\theta)f(\theta)}{1-F(\theta)}\\\\
    &=\frac{\theta\, \varphi'(\theta)}{\varphi(\theta)}\cdot \omega(\theta)+\frac{\theta\,f(\theta)}{1-F(\theta)}-1
\end{align*}
where the last equality follows from  the definition of $\varphi(\cdot)$.
\end{proof}\bigskip

\noindent\begin{proof}[Proof of \autoref{cor:1}]
Fix $k',k''\in(0,1)$ with $k'<k''$. By \autoref{prop:cutoffcompstat}, 
$\ts(k')<\ts(k'')$. For any buyer type $\theta\in[\ts(k'),\ts(k'')]$, the buyer's payoff difference is given by
\begin{align*}
    U(\theta,k'')-U(\theta,k')&= \theta\beta(k'')-\big(\theta-p^*(k')\big)\\[6pt]
    &=\ts(k')\big(1-\omega(\ts(k'))\big)-\theta\big(1-\omega(\ts(k''))\big)\\[6pt]
    &\geq \ts(k')\big(1-\omega(\ts(k'))\big)-\ts(k'')\big(1-\omega(\ts(k''))\big)\\[6pt]
    &\geq 0,
\end{align*}
where the second equality follows from \eqref{eq:foc_omega} and the definition of $p^*(k)$, the first inequality follows because $\theta\leq\ts(k'')$ and $1-\omega(\ts(k''))\geq 0$, and the last inequality follows because the mapping $\theta\mapsto \theta(1-\omega(\theta))$ is assumed to be non-increasing and $\ts(k')<\ts(k'')$.  Therefore, $U(\theta,k'')\geq U(\theta,k')$ for all $\theta\in[\ts(k'),\ts(k'')]$.

Conversely, suppose the payoff inequality holds for all such $k',k''$ and all $\theta\in[\ts(k'),\ts(k'')]$. Taking $\theta=\ts(k'')$, notice that the payoff inequality implies
\begin{align*}
    0&\leq U(\ts(k''),k'')-U(\ts(k''),k')=\ts(k')\big(1-\omega(\ts(k'))\big)-\ts(k'')\big(1-\omega(\ts(k''))\big).
\end{align*}
Since $k\mapsto \ts(k)$ is continuous and strictly increasing with image $(\theta^M,\bar\theta)$, the above inequality is equivalent to $\theta\mapsto \theta(1-\omega(\theta))$ being non-increasing on $(\theta^M,\bar\theta)$.
\end{proof}\bigskip

\noindent\begin{proof}[Proof of \autoref{prop:highcs}]
For each  $k\in (0,1)$ and $\theta\in \Theta$, 
\[
U(\theta, k)\geq \min\left\{\frac{k}{F(\ts(k))}, 1\right\} \theta\geq k\, \theta,
\]
where the first inequality follows because a buyer's equilibrium payoff is at least as high as her payoff from only relying on the public option, and the last  follows because $\ts(k)>F^{-1}(k)$ and $\theta\geq \underline \theta\geq 0$. Furthermore, since $\ts(k)<\bar \theta$, the last inequality is strict for all $\theta>0$. 
\end{proof}\bigskip

\noindent\begin{proof}[Proof of \autoref{prop:cournot}]
Suppose there exists a symmetric equilibrium of the $N$-firm Cournot oligopoly in which firms earn strictly positive profits. Consider such an equilibrium in which each firm supplies quantity $q^*>0$ with $P(Nq^*)q^*>0$. Since $P(Q)=0$ whenever $ Q \geq 1- k$, it must be that $Nq^*<1-k$. Moreover, from the firm's first-order condition, the equilibrium quantity $q^*$ solves 
\[
P(Nq^*)+P'(Nq^*)q^*=0. 
\]

Recall that each aggregate quantity $Q$ induces a cutoff type $\tc(Q)$ such that $1-F(\tc(Q))=Q$. Since the inverse demand function, the aggregate equilibrium quantity, and each firm's symmetric equilibrium supply can all be expressed as functions of the induced cutoff type,  we can rewrite the above first-order condition exclusively as a function of the equilibrium cutoff $\theta^*_N\coloneqq \tc(Nq^*)$:
\begin{align*}
r_N(\theta^*_N)\coloneqq \frac{k}{F(\theta^*_N)^2}\left[\frac{1}{N}\theta^*_N\,\big(1-F(\theta^*_N)\big)+\varphi_N(\theta^*_N)F(\theta^*_N)\right]-\varphi_N(\theta^*_N)=0.
\end{align*} 
Notice that  $r_N(\theta^*_N)=0$ is equivalent to $\theta^*_N$ solving \eqref{eq:cournot_foc}.

Recall $r(\cdot)$ from the proof of \autoref{prop:optimal}. When $N=1$, then $r_N\equiv r$, and thus \eqref{eq:cournot_foc} is a generalization of the monopoly's first-order condition \eqref{eq:foc} to the oligopoly setting. Furthermore, 
\begin{align*}
    N\cdot r_N(\theta)-r(\theta)=(N-1)\,\theta\left(\frac{k}{F(\theta)}-1\right).
\end{align*}
When $N\geq 2$, we have $N\cdot r_N(\theta)>r(\theta)$ for all $0<\theta<F^{-1}(k)$ and $N\cdot r_N(\theta)<r(\theta)$ for all $\theta>F^{-1}(k)$. In particular, the two functions coincide at $\theta=F^{-1}(k)$. Moreover, the mapping $\theta\mapsto (N\cdot r_N(\theta)-r(\theta))$ is strictly decreasing in $\theta$ over the interval $(F^{-1}(k),\bar \theta)$.

Since $r_N$ is continuous, establishing the existence of a unique symmetric equilibrium with positive profits is equivalent to showing that \eqref{eq:cournot_foc} has a unique solution $\theta^*_N\in(F^{-1}(k), \bar\theta)$.

To that end, notice that for all $\theta\leq F^{-1}(k)$, we have $N\cdot r_N(\theta)\geq r(\theta)>0$, where the last inequality was established in the proof of \autoref{prop:optimal}. Thus, there is no solution to \eqref{eq:cournot_foc} on the interval $[\underline\theta, F^{-1}(k)]$.

Next, consider any $F^{-1}(k)<x<y<\bar\theta$. In this case, 
\begin{align*}
N\big(r_N(y)-r_N(x)\big)\leq r(y)-r(x)<0
\end{align*}
where the first inequality follows because $N\cdot r_N(\theta)-r(\theta)$ is non-increasing in $\theta$ for all $N\geq 1$, and the second inequality follows because $r$ is strictly decreasing on $(F^{-1}(k),\bar\theta)$, as established in the proof of \autoref{prop:optimal}. Moreover, 
\[
N\cdot r_N(F^{-1}(k))=r(F^{-1}(k))>0>r(\bar\theta)>N\cdot r_N(\bar\theta),
\]
where the first and second inequalities were established in the proof of \autoref{prop:optimal}. Hence, there is a unique $\theta^*_N\in (F^{-1}(k), \bar\theta)$ that solves \eqref{eq:cournot_foc}. We have thus established that $\theta^*_N$ is the unique solution to \eqref{eq:cournot_foc} over $\Theta$, and consequently, characterizes the unique symmetric equilibrium in which firms earn a positive profit. 

Next, let us show $\theta^*_N>\theta^O_N$. This is trivially true if $\theta^O_N\leq F^{-1}(k)$, so let us focus instead on the case that $\theta^O_N> F^{-1}(k)$. Since $F^{-1}(k)>\underline \theta$ for $k>0$, this implies that $\theta^O_N>\underline \theta$. This in turn implies $\varphi_N(\theta^O_N)=0$. Since the right-hand side of \eqref{eq:cournot_foc} is positive for any $\theta\in\Theta$, we have $\varphi_N(\theta^*_N)>0$. Hence, by strict monotonicity of $\varphi_N$, we conclude that $\theta^*_N> \theta^O_N$.

Next, consider any $N''>N'\geq 1$ and their corresponding optimal cutoffs $\theta^*_{N''}$ and $\theta^*_{N'}$, respectively.  Over the interval $(F^{-1}(k), \bar\theta)$, both $r_{N''}$ and $r_{N'}$ are strictly decreasing functions with $r_{N''}(\theta)<r_{N'}(\theta)$. Since $r_{N''}(\theta^*_{N''})=r_{N'}(\theta^*_{N'})=0$,  we must have $\theta^*_{N''}<\theta^*_{N'}$. In particular, this establishes that for any $N>1$, $\theta^*_N<\ts$. 

Finally, since $\theta^*_N\geq F^{-1}(k)$ for each $N\geq 1$, we have $\lim_{N\to\infty}\theta^*_N\eqqcolon\theta^*_\infty\geq F^{-1}(k)$. For the sake of contradiction, suppose $\theta^*_\infty> F^{-1}(k)$. This implies that for any $\theta\in(F^{-1}(k), \theta^*_\infty)$ and any $N$, $r_N(\theta)>0$. Consequently, for any such $\theta$, we have $\lim_{N\to\infty}r_N(\theta)\geq 0$. However, for any $\theta>F^{-1}(k)$,
\[
\lim_{N\to\infty} r_N(\theta)=\theta\left(\frac{k}{F(\theta)}-1\right)<0,
\]
yielding a contradiction. As a result, $\theta^*_\infty=F^{-1}(k)$.
\end{proof}\bigskip

\noindent\begin{proof}[Proof of \autoref{prop:cournot_comparative}]

\noindent \emph{Part $(a)$}:  For a given type $\theta\in \Theta$ and capacity $k\in (0,1)$, define
\[
r_N(\theta,k)\coloneqq \frac{k}{F(\theta)^2}\left[\frac{1}{N}\theta\big(1-F(\theta)\big)+\varphi_N(\theta)F(\theta)\right]-\varphi_N(\theta),
\]
which recall (from the proof of \autoref{prop:cournot}) is the derivative of the profit function for a given marginal type $\theta> F^{-1}(k)$. Notice that $r_N(\theta, k)$ is continuous in both arguments. Moreover, for each $k\in (0,1)$, the optimal cutoff type $\theta^*_N(k)$ satisfies $r_N(\theta^*_N(k), k)=0$. Since $\theta^*_N(k)$ is the unique such solution, we immediately obtain the continuity of the mapping $k\mapsto \theta^*_N(k)$ over $(0,1)$.

For each $k$, the mapping $\theta\mapsto r_N(\theta, k)$ is strictly decreasing over $(F^{-1}(k), \bar \theta)$, as established in the proof of \autoref{prop:cournot}. Furthermore, for each $\theta\in (\underline \theta, \bar \theta)$, 
\begin{align*}
\frac{\partial r_N(\theta, k)}{\partial k}& =\frac{1}{F(\theta)^2}\left[\frac{1}{N}\theta\big(1-F(\theta)\big)+\varphi_N(\theta)F(\theta)\right]\\[6pt]
&=\frac{1}{F(\theta)^2}\left[\frac{1}{N}\theta\big(1-F(\theta)\big)+\frac{1}{N}\varphi(\theta)F(\theta)+ \frac{N-1}{N}\theta F(\theta)\right]\\[6pt]
&=\frac{1}{N}\frac{\partial r(\theta, k)}{\partial k}+\frac{1}{F(\theta)}\frac{N-1}{N}\theta\\[6pt]
&>0
\end{align*}
where the inequality follows because $\partial r(\theta, k)/\partial k>0$ for all $\theta\in(\underline\theta, \bar\theta)$ as established in the proof of \autoref{prop:cutoffcompstat}. Thus, for any $k',k''\in (0,1)$ with $k''>k'$, 
\[
0=r_N(\theta^*_N(k'), k')=r_N(\theta^*_N(k''), k'')>r_N(\theta^*_N(k''), k'),
\]
which implies that $\theta^*_N(k'')>\theta^*_N(k')$, as desired.

Let $\theta^*_N(1)\coloneqq \lim_{k\to 1}\theta^*_N(k)$ and $\theta^*_N(0)\coloneqq \lim_{k\to 0}\theta^*_N(k)$. To establish the first limit result, note that $\theta^*_N(k)\in (F^{-1}(k), \bar \theta)$ for all $k\in (0,1)$. Thus, $\bar \theta=\lim_{k\to 1}F^{-1}(k)\leq\theta^*_N(1)\leq \bar \theta$, so we conclude that $\bar\theta=\theta^*_N(1)$, as desired.

To establish the second limit result, note that $\theta^*_N(k) \geq \theta^O_N$ for all $k \in (0,1)$, which implies $\theta^*_N(0) \geq \theta^O_N$. Suppose, for the sake of contradiction, that $\theta^*_N(0) > \theta^O_N$. Thus, $\theta^*_N(0) > \underline{\theta}$ and $F(\theta^*_N(0)) > 0$. As a result, $(\theta,k)\mapsto r_N(\theta, k)$ is continuous at the point $(\theta^*_N(0), 0)$. Since $r_N(\theta^*_N(k), k)=0$ for all $k \in (0,1)$, taking the limit as $k \to 0$, we obtain $r_N(\theta^*_N(0),0)=0$. On the other hand, by definition, $r_N(\theta^*_N(0),0)=-\varphi_N(\theta^*_N(0))<0$, where the inequality follows from the strict monotonicity of $\varphi_N$ and the assumption that $\theta^O_N<\theta^*_N(0)$, yielding a contradiction. Thus, it must be that $\theta^*_N(0)=\theta^O_N$, as desired. \bigskip

\noindent \emph{Part $(b)$}: From \eqref{eq:foc_omega_cournot}, we have that $\beta_N(k)=\omega_N(\theta_N^*(k))$ for all $k\in (0,1)$. The proof for Part $(b)$ of the proposition then follows identically to the proof of \autoref{prop:lowvaluecompstat}, with $\omega_N(\cdot)$ and $\theta^*_N(k)$ replacing $\omega(\cdot)$ and $\ts(k)$, respectively.\bigskip

\noindent \emph{Part $(c)$}: Since $\beta_N(k)=\omega_N(\theta_N^*(k))$ for all $k\in (0,1)$, we can express the Cournot equilibrium price as $p^*_N(k)=\theta^*_N(k)\big(1-\omega_N(\theta^*_N(k))\big)$. 

Fix any $k',k''\in(0,1)$ with $k'<k''$. The proof for Part $(c)$ of the proposition for the case that $\theta>\theta^*_N(k'')$ then follows identically to the proof of \autoref{prop:highvaluecscompstat}, with $\omega_N(\cdot)$ and $\theta^*_N(k)$ replacing $\omega(\cdot)$ and $\ts(k)$, respectively. The proof for Part $(c)$ for the case that $\theta\in[\theta_N^*(k'),\theta^*_N(k'')]$ similarly follows identically to the proof of \autoref{cor:1}.
\end{proof}\bigskip

\noindent\begin{proof}[Proof of \autoref{prop:complement_vs_baseline}]
For any  $k\in (0,1)$, $\ts(k)<\bar\theta$ (by \autoref{prop:optimal}). Hence, $\beta(k)=k/F(\ts(k))>k$, which implies that 
\[
U(\theta, k)-\widetilde U(\theta, k)=\theta\big(\beta(k)-k\big)>0
\]
for all always-low-value types $\theta\leq \theta^M$.

Next, notice that for any switcher $\theta\in(\theta^M,\ts(k))$, 
\begin{align*}
   U(\theta,k)-\widetilde U(\theta,k)=\theta^M(1-k)-\theta\big(1-\beta(k)\big)
\end{align*}
is strictly decreasing in $\theta$, while for any always-high-value type $\theta\geq\ts(k)$, 
\[
U(\theta, k)-\widetilde U(\theta, k)=\theta^M(1-k)-p^*(k)
\]
is a constant. Thus, if $U(\ts(k),k)\geq \widetilde U(\ts(k), k)$, then $U(\theta,k)\geq \widetilde U(\theta,k)$ for all $\theta\in\Theta$. Alternatively, if  $U(\ts(k),k)<\widetilde U(\ts(k), k)$, then by continuity of the payoffs, there exists some type $\hat\theta(k)\in (\theta^M, \ts(k))$ such that $U(\hat\theta(k),k)=\widetilde U(\hat\theta(k), k)$. Moreover, $U(\theta,k)>\widetilde U(\theta,k)$ for all $\theta<\hat\theta(k)$ and $U(\theta,k)<\widetilde U(\theta,k)$ for all $\theta>\hat\theta(k)$.

Hence, $U(\theta,k)\geq \widetilde U(\theta,k)$ for all $\theta\in\Theta$ and all $k\in (0,1)$ if $U(\ts(k),k)\geq \widetilde U(\ts(k), k)$ for all $k\in (0,1)$. Notice that
\begin{align*}
U(\ts(k),k)-\widetilde U(\ts(k), k)&= \theta^M(1-k)-p^*(k)\\[6pt]
&=\theta^M\Big(1-\omega\big(\ts(k)\big)F\big(\ts(k)\big)\Big)-\ts(k)\Big(1-\omega\big(\ts(k)\big)\Big),
\end{align*}
where the equality follows from \eqref{eq:foc_omega}. Since $\ts(k)$ is continuous in $k$ with $\lim_{k\to 0}\ts(k)=\theta^M$ and $\lim_{k\to 1}\ts(k)=\bar\theta$, we have that $U(\ts(k),k)\geq \widetilde U(\ts(k), k)$ for all $k\in (0,1)$ if and only if $\theta^M\big(1-\omega(\theta)F(\theta)\big)\geq \theta\big(1-\omega(\theta)\big)$ for all $\theta\in (\theta^M,\bar\theta)$. By rearranging the terms, we obtain \eqref{eq:complement condition}.

Alternatively, if \eqref{eq:complement condition} fails for some $\theta'\in(\theta^M,\bar\theta)$, then there exists some $k'\in(0,1)$ such that $\theta'=\ts(k')$ and  $U(\ts(k'),k')-\widetilde U(\ts(k'), k')<0$. Because the mapping $k\mapsto U(\ts(k),k)-\widetilde U(\ts(k), k)$ is continuous over $(0,1)$, there exists an open set $K\subseteq (0,1)$ such that for all $k\in K$, we have $U(\ts(k),k)-\widetilde U(\ts(k), k)<0$, which yields the desired converse result. 
\end{proof}

\subsection{Optimality of a posted-price mechanism}\label{sec:post}
This section establishes that the monopolist’s profit-maximizing mechanism when facing a public option is a posted price. Accordingly, the restriction to posted-price mechanisms in the main text is without loss of generality. I establish this result in a setting that generalizes the baseline model in two directions: the public option may supply a lower-quality good than the monopolist, and it may charge a subsidized (possibly zero) price.

Formally, the monopolist supplies a good whose quality is normalized to 1, and seeks to maximize its profit. The public option offers a good of quality $\delta\in (0,1]$ at an exogenously fixed price of $\rho\geq 0$. A buyer with valuation $\theta\in\Theta$ who pays $\mathrm{t}\geq 0$ and receives a good of quality $\tilde \delta\in \{\delta, 1\}$ with probability $\mathrm{x}\in[0,1]$ earns a payoff of $\mathrm{x} \theta \tilde \delta -\mathrm{t}$. Observe that we recover the baseline model when $\delta=1$ and $\rho=0$. Finally, recall that if the demand at the public option is $d$, then its allocation probability is $\min\{k/d,1\}$, with $\min\{k/0,1\}=1$. 

It is clear that any buyer type $\theta<\underline s\coloneqq \max\{\underline\theta, \rho/\delta\}$ would never use the public option. If $\rho \geq \delta  \theta^M$, then the public option would be irrelevant altogether; the monopolist would continue to charge $\theta^M$ and no buyer type would opt for the public option. I therefore assume $\rho < \delta \theta^M$.

I consider direct revelation mechanisms of the form $(x,t)\in\mathcal{X}\times\mathcal{T}$, where a buyer who reports type $\hat \theta\in\Theta$ pays $t(\hat \theta)$ to the monopolist and is allocated a good from the monopolist with probability $x(\hat \theta)$. With the complementary probability $1-x(\hat \theta)$, the buyer can either rely on the public option or exit the market without receiving the good.

If all buyers report their types truthfully, the public option's induced demand is 
\begin{equation*}
    d(x)\coloneqq \int_{\underline s}^{\bar\theta} \big(1-x(\theta)\big)dF(\theta),
\end{equation*}
with the maximal public option demand given by $\bar d\coloneqq 1-F(\underline s)$. Accordingly, a type $\theta$ who reports $\hat \theta$ while all other buyers report truthfully earns an expected payoff of
\begin{equation*}
    U(\hat \theta,\theta|x,t)\coloneqq x(\hat \theta) \theta+\big(1-x(\hat \theta)\big)\cdot \min\left\{1,\frac{k}{d(x)}\right\} (\delta\theta-\rho)^+-t(\hat \theta),
\end{equation*}
where for any scalar $z\in\mathbb{R}$,  $z^+\coloneqq \max\{z,0\}$.

A mechanism $(x,t)$ is incentive compatible over a subset $S\subseteq \Theta$ if 
\begin{equation}
    \label{eq:oaic}
    \tag{IC$_S$}
    U(\theta,\theta|x,t)\geq U(\hat \theta,\theta|x,t),\; \forall \theta,\hat \theta\in S,
\end{equation}
and the mechanism is individually rational over $S\subseteq \Theta$ if 
\begin{equation}
    \label{eq:oair}
    \tag{IR$_S$}
    U(\theta,\theta|x,t)\geq \min\left\{1,\frac{k}{d(x)}\right\}(\delta\theta-\rho)^+, \; \forall \theta\in S.
\end{equation}
The monopolist's objective is to maximize its expected revenue by offering a mechanism that is incentive compatible and individually rational over $\Theta$. 

Before solving for the optimal mechanism, notice that any buyer type $\theta<\rho/\delta$ would never use the public option. Since I already assumed that $\theta^M>\rho/\delta$, it is optimal for the monopolist to exclude any buyer type $\theta<\rho/\delta$ from the private market as well, i.e., $x(\theta)=t(\theta)=0$ for all $\theta<\rho/\delta$.

Recall $\underline s= \max\{\underline \theta, \rho/\delta\}$. Henceforth, consider the subset of types $S=[\underline s, \bar\theta]\subseteq\Theta$. I first analyze the following relaxed problem:
\begin{align*}
\label{eq:oamd}
\tag{MD}
    \max_{(x, t)\in \mathcal{X}\times \mathcal{T}} \hspace*{.2em} &\int_S t(\theta)dF(\theta) \hspace{.2em} \text{ s.t. }  (x,t)\hspace{.2em}  \text{ satisfies }  \eqref{eq:oaic} \text{ and } \eqref{eq:oair}.
\end{align*}
I shall ignore types $\theta\notin S$ for now and show at the end that the solution to the relaxed problem \eqref{eq:oamd} is in fact incentive compatible and individually rational over $\Theta$, i.e., $(i)$ no buyer type $\theta<\rho/\delta$ benefits from misreporting as a higher type $\hat\theta\geq \rho/\delta$, and $(ii)$ no buyer type $\theta\geq \rho/\delta$ benefits from misreporting as a lower type $\hat\theta<\rho/\delta$.

For any given mechanism $(x,t)$ and any type $\hat\theta\in S$, define the \emph{transformed allocation rule} by  
\[
\chi(\hat \theta|x)\coloneqq
    x(\hat \theta)\left(1-\delta\cdot \min\left\{\frac{k}{d(x)},1\right\}\right),
\]
and define the \emph{transformed transfer rule} by  
\[
\tau(\hat \theta|x, t)\coloneqq t(\hat \theta)-\rho x(\hat \theta) \cdot \min\left\{\frac{k}{d(x)},1\right\}.
\]
If a buyer with valuation $\theta\in S$ reports a type $\hat\theta\in S$, then her payoff net of the outside option, which I denote by $u(\theta|x,t)$, can be written as
\begin{equation*}
\begin{split}
 U(\hat \theta, \theta|x,t)-(\delta\theta-\rho)\cdot \min\left\{\frac{k}{d(x)}, 1\right\}=\chi(\hat \theta|x)\theta-\tau(\hat \theta|x,t).
\end{split}
\end{equation*}
Thus, a mechanism $(x,t)$ satisfies \eqref{eq:oaic} if and only if  
\begin{equation}
    \label{eq:oaic2}
    \tag{IC$'_S$}
    \chi(\theta|x)\theta-\tau(\theta|x,t)\geq \chi(\hat \theta|x)\theta-\tau(\hat \theta|x,t), \hspace{.5em} \forall \theta,\hat \theta\in S.
\end{equation}
Similarly, $(x,t)$ satisfies \eqref{eq:oair} if and only if  
\begin{equation}
    \label{eq:oair2}
    \tag{IR$'_S$}
    \chi(\theta|x)\theta-\tau(\theta|x,t)\geq 0, \hspace{.5em} \forall \theta\in S.
\end{equation}

Hence, leveraging \cite{myerson1981optimal}, a mechanism $(x,t)$ satisfies \eqref{eq:oaic} if and only if
\begin{enumerate}[$(a)$]
    \item $\chi(\cdot|x)$ is non-decreasing over $S$, and
    \item For all $\theta\in S$,
\begin{equation*}
    t(\theta)=\chi(\theta|x)\theta+\rho x(\theta)\cdot \min\left\{\frac{k}{d(x)}, 1\right\}-\int_{\underline s}^\theta \chi(s|x)ds-u(\underline s|x,t).
\end{equation*}
\end{enumerate}
Moreover, an incentive-compatible mechanism $(x,t)$ satisfies \eqref{eq:oair} if $u(\underline s|x,t)\geq 0$. Consequently, the monopolist's revenue from an incentive-compatible mechanism $(x,t)$ can be expressed as 
\begin{align*}
&\int_S t(\theta)dF(\theta)\\[6pt]
=&\int_S \chi(\theta|x)\varphi(\theta)dF(\theta)+\rho\cdot \min\left\{\frac{k}{d(x)}, 1\right\}\int_S x(\theta)dF(\theta)-u(\underline s|x,t)(1-F(\underline s))\\[8pt]
= &\int_S x(\theta)\left[\varphi(\theta)\left(1-\delta\cdot  \min\left\{\frac{k}{d(x)}, 1\right\}\right)+\rho\cdot \min\left\{ \frac{k}{d(x)}, 1\right\}\right]dF(\theta)-u( \underline s|x,t)(1-F(\underline s)).   
\end{align*}
Clearly, if $(x,t)$ solves \eqref{eq:oamd}, then $u(\underline s|x,t)=0$. 

Given an allocation rule $x\in\mathcal{X}$ and an arbitrary induced demand $\mathrm{d}\in [0,\bar d]$, define  
\[
R(x,\mathrm{d})\coloneqq \int_S x(\theta)\left[\varphi(\theta)\left(1-\delta\cdot  \min\left\{\frac{k}{\mathrm{d}}, 1\right\}\right)+\rho\cdot \min\left\{ \frac{k}{\mathrm{d}}, 1\right\}\right]dF(\theta)
\]
Notice that \eqref{eq:oamd} can now be equivalently restated as 
\begin{equation}
\label{eq:oamd3}
    \tag{MD$'$}
\max_{x\in \mathcal{X}} \hspace*{.2em} R(x, d(x)) \hspace{.2em} \text{ s.t. }  \chi(\cdot|x) \text{ is non-decreasing}.    
\end{equation}
Unlike the classical mechanism design problem, \eqref{eq:oamd3} is a non-linear optimization problem. However, we can make the problem more tractable by decomposing it into the following  equivalent nested problem:
\[
\max_{\mathrm{d}\in [0, \bar d]}\left\{\max_{x\in \mathcal{X}} R(x,\mathrm{d}) ~ \text{ s.t.} ~~  \chi(\cdot|x) \text{ is non-decreasing over $S$, and } d(x)=\mathrm{d} \right\}.
\]
Let us first solve the inner constrained linear programming problem. To that end, fix some $\mathrm{d}\in [0,\bar d]$ and consider two cases.

\noindent \textbf{Case 1:} Suppose $1-\delta\cdot  \min\{{k}/{\mathrm{d}}, 1\}>0$.

In this case, $\chi(\cdot|x)$ is non-decreasing if and only if $x$ is non-decreasing. 
 Let $\mathcal{X}_{M}$ be the set of non-decreasing allocation functions, which is a convex and compact subset of the set of integrable allocation functions. Define the subset
\[
\mathcal{X}_M^\mathrm{d}\coloneqq\{x\in\mathcal{X}_M: d(x)=\mathrm{d} \}.
\]
Let $\theta^\mathrm{d}$ be the unique value of $\theta\in S$ such that $F(\theta)-F(\underline s)=\mathrm{d}$. Define the allocation rule $x^\mathrm{d}(\theta)\coloneqq \mathbbm{1}[\theta\geq \theta^\mathrm{d}]$, and notice that $x^\mathrm{d}\in\mathcal{X}_M^\mathrm{d}$, so $\mathcal{X}^\mathrm{d}_M$ is non-empty. Moreover, since the mapping $x\mapsto d(x)$ is linear and continuous, the subset $\mathcal{X}_M^\mathrm{d}$ is also a convex and compact set. Therefore, the inner linear programming problem given by
\begin{align*}
\tag{IP}
    \max_{x\in \mathcal{X}_M^\mathrm{d}} R(x,\mathrm{d})
\end{align*}
attains its maximum at an extreme point of $\mathcal{X}_M^\mathrm{d}$. From \cite{winkler1988extreme} (Proposition 2.1), $x$ is an extreme point of $\mathcal{X}_M^\mathrm{d}$ if there exists a weight $\alpha\in[0,1]$ and step functions  $x_1(\theta)=\mathbbm{1}[\theta\geq \theta_1]$ and $x_2(\theta)=\mathbbm{1}[\theta\geq \theta_2]$ with cutoffs $\theta_1,\theta_2\in S$ such that $x=\alpha x_1+(1-\alpha)x_2$ and $d(x)=\mathrm{d}$.

Suppose, for the sake of a contradiction, that the inner problem attains its maximum at $x^*=\alpha x_1+(1-\alpha) x_2$ where $\alpha\in (0,1)$ and the step functions $x_1, x_2$ have cutoffs $\theta_1<\theta_2$, respectively. Since $d(\cdot)$ is linear, we have 
\[
d(x^*)=\alpha d(x_1)+ (1-\alpha) d(x_2)=\alpha F(\theta_1)+ (1-\alpha) F(\theta_2)-F(\underline s).
\]
At the same time,  the fact that $x^*\in\mathcal{X}_M^\mathrm{d}$ implies that $d(x^*)=\mathrm{d}=F(\theta^\mathrm{d})-F(\underline s)$. Equating the two expressions for $d(x^*)$, we have that $\theta^\mathrm{d}\in (\theta_1, \theta_2)$ and 
\[
\alpha=\frac{F(\theta_2)-F(\theta^\mathrm{d})}{F(\theta_2)-F(\theta_1)}.
\]

The monopolist's profit from implementing $x^*$ is then given by
\begin{align*}
R(x^*, \mathrm{d})=&\left(1-\delta\cdot  \min\left\{\frac{k}{\mathrm{d}}, 1\right\}\right)\left[\int^{\bar \theta}_{\theta_2}\varphi(\theta)dF(\theta)+\alpha\int^{\theta_2}_{\theta_1}\varphi(\theta)dF(\theta)\right]\\[6pt]
&+\rho\cdot \min\left\{ \frac{k}{\mathrm{d}}, 1\right\}\Bigg[1-\underbrace{\Big(\alpha F(\theta_1)+(1-\alpha)F(\theta_2)\Big)}_{=F(\theta^\mathrm{d})}\Bigg]\\[6pt]
< &\left(1-\delta\cdot  \min\left\{\frac{k}{\mathrm{d}}, 1\right\}\right)\int^{\bar \theta}_{\theta^\mathrm{d}}\varphi(\theta)dF(\theta)+\rho(1-F(\theta^{\mathrm{d}}))\cdot \min\left\{ \frac{k}{\mathrm{d}}, 1\right\}\\[6pt]
=&R(x^\mathrm{d}, \mathrm{d}),
\end{align*}
where the inequality follows because the strict monotonicity of $\varphi$ implies 
\[
\mathbb{E}_F\big[\varphi\big|\theta\in [\theta_1, \theta_2]\big]<\mathbb{E}_F\big[\varphi\big|\theta\in [\theta^\mathrm{d}, \theta_2]\big].
\]
Consequently, $x^\mathrm{d}$ is the essentially unique allocation rule in $\mathcal{X}_M^\mathrm{d}$ at which the inner problem attains its maximum, and the value function for the inner problem is given by $R(x^\mathrm{d}, \mathrm{d})$.
\medskip

\noindent \textbf{Case 2:} Suppose $1-\delta\cdot  \min\{{k}/{\mathrm{d}}, 1\}= 0$.

In this case, $\chi(\theta|x)=0$ for all $\theta\in S$ and all $x\in\mathcal{X}$. Hence, the optimal solution is any allocation rule $x\in \mathcal{X}$ (monotone or not) such that $d(x)=\mathrm{d}$. In particular, $x^\mathrm{d}$ is a solution, and the value function for the inner problem is once again given by $R(x^\mathrm{d}, \mathrm{d})$.
\medskip

Let us next solve the outer optimization problem:
\[
\max_{\mathrm{d}\in [0,\bar d]} R(x^\mathrm{d}, \mathrm{d}).
\]
Since $\mathrm{d}=F(\theta^\mathrm{d})-F(\underline s)$ by construction, choosing $\mathrm{d}\in [0,\bar d]$ is equivalent to choosing a cutoff type $\theta\in S$. Hence, the outer problem is equivalent to:
\begin{align*}\label{eq:oaop}
     \max_{\theta\in S}\, (1-F(\theta))\left[\theta-\min\left\{\frac{k}{F(\theta)-F(\underline s)}, 1\right\}\cdot(\delta\theta-\rho)\right],
 \tag{OP}
\end{align*}
with the maximum attained at some $\ts\in S$. The optimal mechanism is then given by a posted price mechanism $(x^*,t^*)$ with 
\[
x^*(\theta)=\left\{ \begin{array}{ccc}
   0 & \mbox{if} & \theta<\ts  \\
   1 & \mbox{if} & \theta\geq \ts  \\
\end{array}\right.
\]
and 
\[
t^*(\theta)=\left\{ \begin{array}{ccc}
   0 & \mbox{if} & \theta<\ts  \\
   \ts-\min\left\{\frac{k}{F(\ts)-F(\underline s)}, 1\right\}\cdot(\delta\ts-\rho) & \mbox{if} & \theta\geq \ts  \\
\end{array}\right. .
\]
Furthermore, the maximizer $\ts$ is generically unique, which implies that there is a unique posted price that solves the monopolist's profit maximization problem. 

Finally, let us confirm that the derived mechanism $(x^*,t^*)$ is incentive compatible and individually rational over $\Theta$. 
For any type $\theta\in S$, the payoff from misreporting as type $\hat\theta\notin S$ would be
\[
U(\hat\theta,\theta|x^*,t^*)=(\delta\theta-\rho)\min\left\{\frac{k}{F(\ts)-F(\underline s)}, 1\right\}\leq U(\theta, \theta|x^*,t^*), 
\]
and thus, all buyer types $\theta\in S$ have an incentive to report truthfully. Similarly, for any type $\theta\notin S$, the payoff from misreporting as type $\hat\theta\in S$ would be
\[
U(\hat\theta,\theta|x^*,t^*)=0=U(\theta, \theta|x^*,t^*)
\]
if $\hat\theta<\ts$, and 
\[
U(\hat\theta,\theta|x^*,t^*)\leq (\delta\theta-\rho)\min\left\{\frac{k}{F(\ts)-F(\underline s)}, 1\right\}<0= U(\theta, \theta|x^*,t^*)
\]
if $\hat\theta\geq \ts$. Thus, all buyer types $\theta\notin S$ also have an incentive to report truthfully. It is trivial to check that $U(\theta,\theta|x^*,t^*)\geq 0$ for all $\theta\in\Theta$. Therefore, $(x^*,t^*)$ is an optimal mechanism that solves the non-relaxed problem. 

\subsubsection*{Discussion.}

As $\delta\to 1$ and $\rho\to 0$, the optimal cutoff that solves \eqref{eq:oaop} is close to the optimal cutoff that solves \eqref{eq:foc}. Hence, the equilibrium characterization in \autoref{prop:optimal} along with the associated comparative statics of the baseline model (\autoref{prop:cutoffcompstat}-\autoref{prop:highvaluecscompstat}) extend to the general setting when $\delta$ is close to 1 and $\rho$ is close to zero.

However, this general setting also allows for other outcomes depending on the parameters of the model. For example, if $k$ is sufficiently large and $\rho$ is sufficiently close to $\delta\theta^M$ (this implies that $\underline s$ is close to $\theta^M$), the optimal cutoff solving \eqref{eq:oaop} is given by
\[
\theta^{**}=\min\{\theta\in\Theta: (1-\delta)\varphi(\theta)+\rho\geq 0\}.
 \]
In this case, the monopolist no longer finds it worthwhile to congest the public option and induce rationing. Instead, it is more profitable for the monopolist to ``undercut'' the public option's price $\rho$. Doing so does not require a significant drop from the standard monopoly price, and roughly the same buyer types continue to purchase from the private market while few, if any, consumers rely on the public option.

Finally, notice that $\theta^{**}$ would be the optimal cutoff also when $k\to 1$ for any $\rho$ and $\delta$. In the limiting case, the mixed market resembles one of a dominant monopoly competing with a fringe market that supplies an inferior good of quality $\delta$ at a constant marginal cost $\rho$. 
\end{appendix}

  \endgroup
\singlespacing
 \bibliographystyle{plainnat}
\nocite{}\bibliography{bibref}
\end{document}